\pdfoutput=1

\documentclass[]{spie}  

\usepackage{graphicx}

\def\etal{{\it et al}.}
\def\micron{{$\mu$m}}

\def\degree{{${}^\circ$}}
\def\kmsq{{km${}^2$}}
\def\msq{{m${}^2$}}
\def\cmsq{{cm${}^2$}}

\def\flux{cm${}^{-2}$s${}^{-1}$sr${}^{-1}$GeV}
\def\HESS{{H.E.S.S.}}
\def\Cherenkov{{\v{C}erenkov}}

\begin{document}

\thispagestyle{empty}
\renewcommand{\thefootnote}{\fnsymbol{footnote}}

\begin{flushright}
{\small
SLAC--PUB--12786\\
Sep, 2007\\}
\end{flushright}

\vspace{.8cm}

\begin{center}
{\bf\large   
Detectors for Cosmic Rays on Ground and in Space\footnote{This work was supported in part by the U.S. Department of Energy under Grant DE-AC02-76SF00515.}}

\vspace{1cm}

Hiroyasu~Tajima
\medskip

Stanford Linear Accelerator Center, Stanford, CA, USA
\end{center}

\vfill

\begin{center}
{\bf\large   
Abstract }
\end{center}

\begin{quote}
The origin of the cosmic rays has been a great mystery since they were discovered by Victor Hess in 1912.
AGASA's observation of ultra-high-energy cosmic-rays (UHECR) possibly beyond the GZK (Greisen, Zatsepin and Kuzmin) cutoff stimulated the field in great deal.
In addition, Kamiokande's detection of neutrinos from SN1987A and the \HESS's detection of TeV gamma-rays from supernova remnants demonstrated the viability of neutrino and TeV gamma-ray astronomy for cosmic-ray research.

A new generation of currently-operating or soon-to-be-operating detectors for charged particles, gamma-rays and neutrinos from cosmos will get us even closer to understanding the nature and origin of cosmic rays.
Detectors for UHECRs, gamma rays and neutrinos are of particular importance in order to study the origins of cosmic rays since these particles are free from the deflection due to magnetic fields.
Detectors for antiparticles and gamma rays would be useful to detect cosmic rays originated from the decay of the dark matter in the Universe.

I will review these cosmic-ray detectors with particular attention on the differences of ground-based, balloon-borne and satellite-borne detectors.
\end{quote}

\bigskip


\vfill

\begin{center} 
{\it Invited Talk at} 
{\it 11th Vienna Conference on Instrumentation \\
Vienna, Austria, February 19--24, 2007} \\
{\it To be published in} 
{\it Nuclear Instruments and Methods A}\\



\end{center}

\clearpage
\pagestyle{plain}

\section{Introduction}
The origin of the cosmic rays has been a great mystery since they were discovered by Victor Hess in 1912.  
Supernova remnants (SNRs) are considered to be the best candidates for hadronic acceleration up to the so-called ``knee" ($3\times10^{15}$ eV) in the cosmic-ray spectrum. 
The evidence to support this was mostly circumstantial, based on theory and logic rather than on observations.
For example, diffusive shock acceleration at SNR shock front could accelerate particles to the desired energies \cite{Blandford87} and the kinetic energy released by supernova explosions can account for the necessary amount of energy \cite{Ginzburg64}.
Synchrotron emission in the X-ray band observed in SN1006 by ASCA \cite{Koyama95} was the first strong indirect evidence for the existence of $\sim$100 TeV electrons in an SNR. 
Observations of TeV gamma rays from an SNR, RX J1713.7-3946, by CANGAROO \cite{Enomoto02} and by \HESS\ \cite{Aharonian04} were the first confirmed direct evidence for a particle accelerated up to 100 TeV in the SNR.
Furthermore, \HESS\ provided the first gamma-ray image of SNR, which is qualitatively consistent with particle acceleration at the shock followed by interaction and radiation in the compressed post-shock region. 
However, the TeV spectrum observed by \HESS\ cannot be interpreted by any single process of leading gamma-ray emission models (inverse Compton, $\pi^0$-decays following proton-proton interactions or non-thermal electron Bremsstrahlung).
Therefore, the cosmic-ray acceleration in the SNR is not yet conclusive.

Above the ``knee" of the cosmic-ray spectrum, there is no obvious candidates although several astronomical objects such as gamma-ray bursts, active galaxies or merging galaxy clusters are suggested as possible origins.
At the high end of the spectrum, cosmic-ray flux is expected to be suppressed above $\sim 10^{19.8}$~eV due to the production of $\Delta$ resonance following the interaction with the cosmic microwave backgrounds (CMB)\cite{GZK}, which is known as GZK suppression.
Above the GZK cutoff energy, cosmic ray cannot travel more than $\sim$20 Mpc.
However, an AGASA measurement\cite{AGASA} indicates continuing spectrum beyond the GZK suppression while a HiRes measurement\cite{HiRes} shows the spectrum consistent with the GZK suppression.
If the AGASA measurement is confirmed, it indicates a local origin of UHECRs.
Observation of UHECRs or UHE-neutrinos associated with known astronomical sources would provide convincing evidences of the UHECR acceleration in these sources.
Note that the phase space of the UHECRs is rather small because charged particles are deflected by the extragalactic and interstellar magnetic field below $\sim 10^{19}$~eV and, at the same time, GZK suppressed above $\sim 10^{19.8}$~eV.
On the other hand, neutrinos are free from both constraints and probe distant Universe.
However, they are difficult to detect due to their small cross-section, which requires extremely large target volume to collect sufficient statistics.
In the case of gamma-ray observations, multi-wavelength analysis is critical to resolve the leptonic and hadronic origin of the gamma-rays. Neutrino detection from the same gamma-ray sources would provide definitive evidences for cosmic-ray acceleration in such sources.

In this paper, I review currently-operating or soon-to-be-operating detectors for charged particles, gamma rays and neutrinos from cosmos on ground or in Space.

\section{Pierre Auger Observatory}
The Pierre Auger Observatory was designed to study cosmic rays at highest energies with high statistics.
The Auger Observatory employs two complementary air-shower detection techniques, a surface array and air fluorescence telescopes.
The surface array measures the lateral structure of the shower at ground level (beyond the shower maximum), with some ability to separate the electromagnetic and muon components. 
On the other hand, the fluorescence detector records the longitudinal profile of the shower development through the atmosphere.

\begin{figure}
\centering
\begin{tabular}{cc}

\parbox{8.2cm}{\centering
\includegraphics[width=7.8cm]{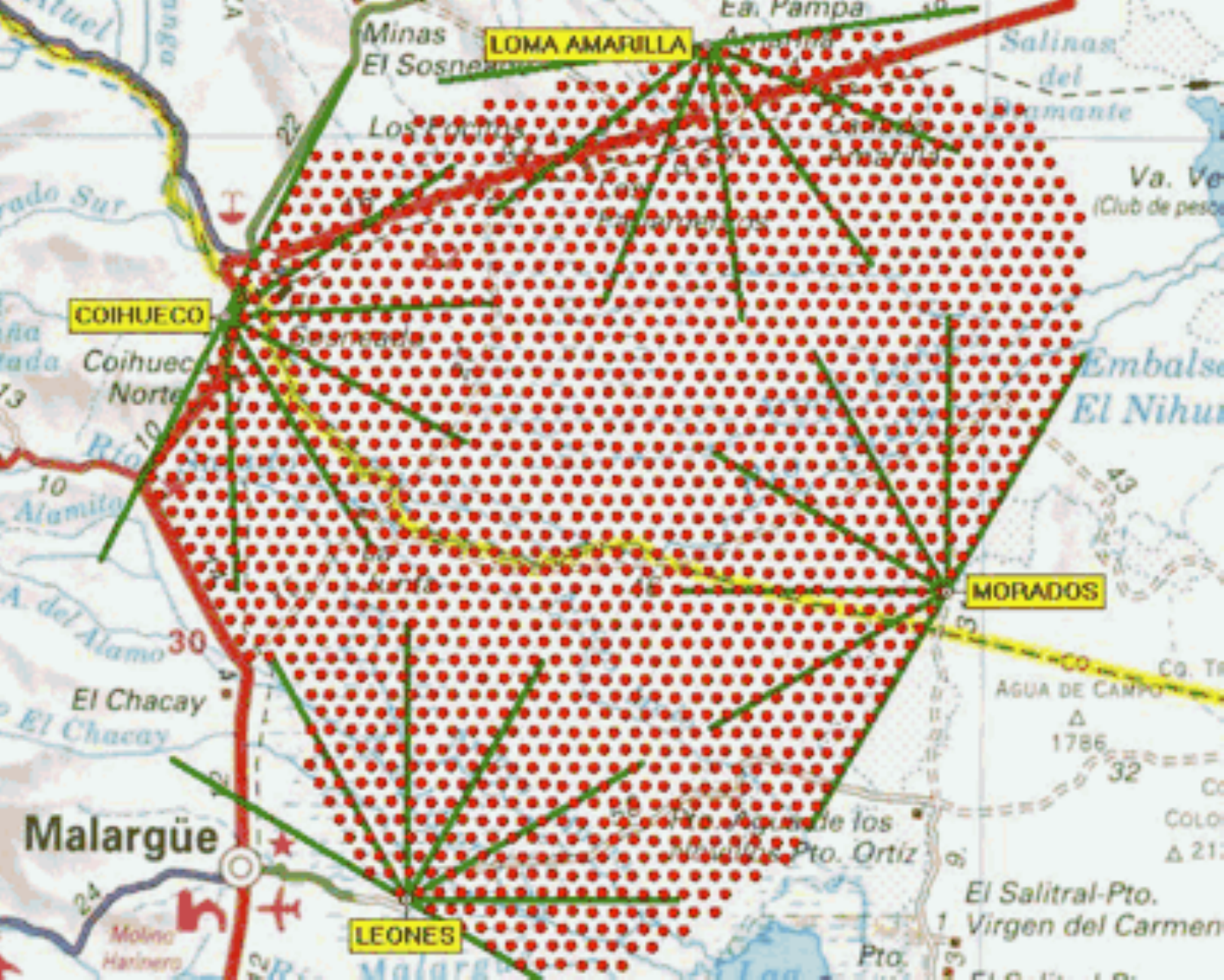}
\caption[The Pierre Auger Observatory plan.]
{\label{fig:PAO-plan}The Pierre Auger Observatory plan.}
}
&
\parbox{8.0cm}{\centering
\includegraphics[width=8.2cm]{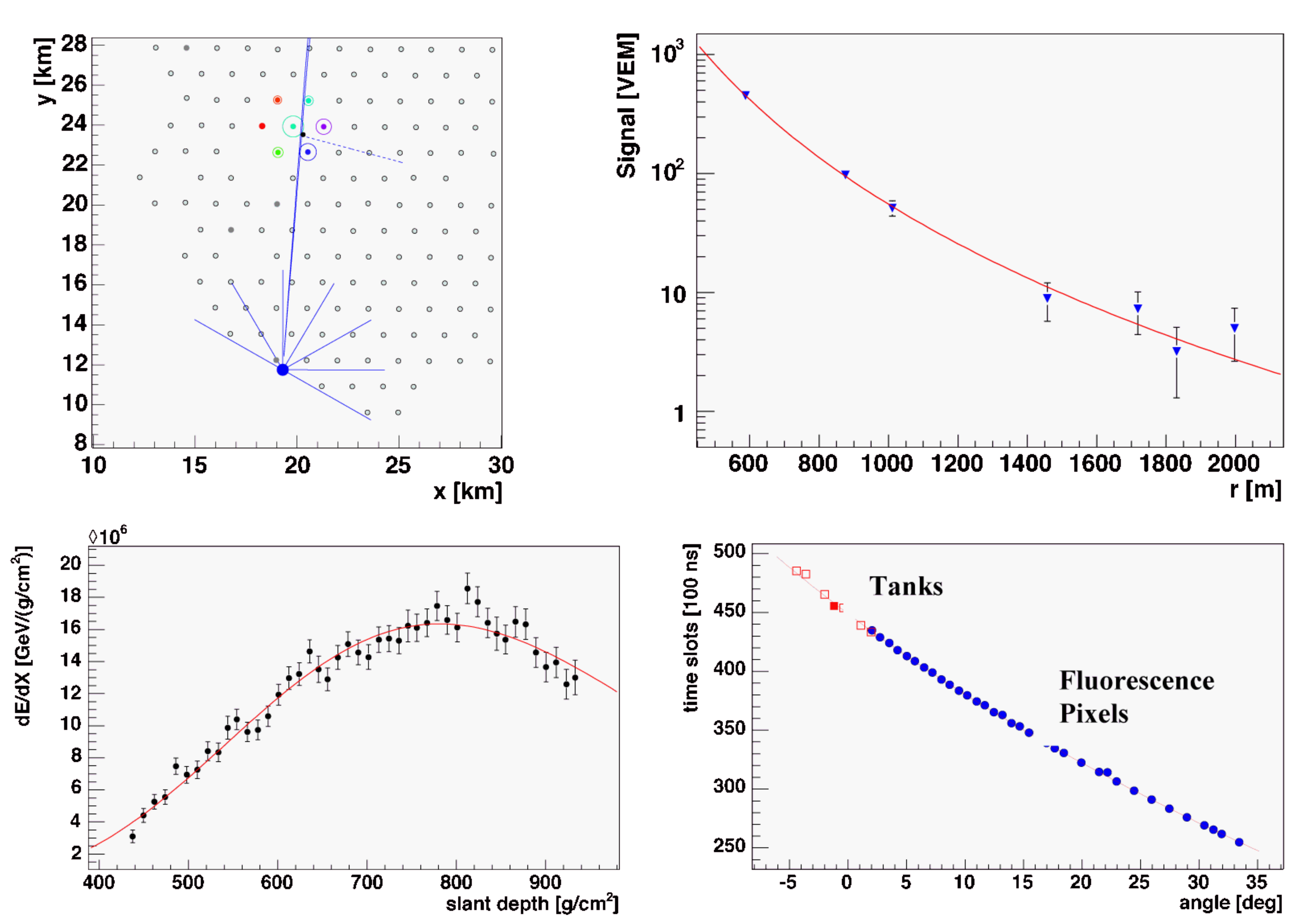}\\
\vspace*{-0.5cm}
\caption[A hybrid event example. Top left panel shows the array hit distribution.
Top right panel shows the lateral density distribution in the surface array for this event.
Bottom left panel shows the energy measurement by the fluorescence detector as a function of the slant depth.
Bottom right panel shows the hit timing as the shower sweeps out the angle $\chi$ formed by the shower and the fluorescence detector.]
{\label{fig:PAO-hybrid}A hybrid event example. Top left panel shows the array hit distribution.
Top right panel shows the lateral density distribution in the surface array for this event.
Bottom left panel shows the energy measurement by the fluorescence detector as a function of the slant depth.
Bottom right panel shows the hit timing as the shower sweeps out the angle $\chi$ formed by the shower and the fluorescence detector.}
}
\end{tabular}
\end{figure}

The Auger Observatory plans to have two sites in the northern and southern hemispheres to cover the entire sky.
Since only the southern site is funded, the northern site is not described in this review.
In order to maximize the aperture for a self-imposed cost limit while ensuring the good efficiency above $10^{19}$~eV, 1,600 surface detectors are being placed on a triangular 1.5~km grid occupying 3000~\kmsq\ as shown in Fig.~\ref{fig:PAO-plan}.
Twenty-four fluorescence telescopes will be placed at the 4 locations around the perimeter of the surface array so that these telescopes view the same region covered by the surface array.

The surface array employs the water \Cherenkov\ particle detector for its robustness and low cost.
The surface detector station is designed to operate 24 hours a day with $\sim$100\% duty cycle in a hash environment with large ambient temperature excursions of 20 degrees in a day, high salinity, dusty air, snow, hail and high humidity inside the tank.
The detector consists of a 12,000 $\ell$ rototionally-molded high density polyethylene water tank containing a sealed laminated polyethylene liner with a reflective inner surface.
The signals from each tank are read out using three 9" photomultipliers (PMTs) and digitized at a sampling frequency of 40~MHz.
A local CPU provides self-triggers and processes events locally.
Timing is obtained by a local GPS unit, and communication to the Central Data Acquisition System is done via a custom built wireless communication system.
Several key parameters are constantly monitored to ensure consistent operation.
Two solar panels provide 10 watts of power for the electronics.
Each detector station is therefore independent and can be operational upon installation, independently of other stations in the array.

A fluorescence site contains six identical fluorescence telescopes.
The telescope design incorporates Schmidt optics, which differentiates the Auger design from that of earlier fluorescence detectors.
The $3.5\times3.5$ \msq\ segmented spherical mirror focuses light from the aperture window on to a camera containing 440 hexagonal 45~mm photomultipliers.
Each PMT covers a 1.5\degree\ diameter portion of the sky while the optical spot size is 0.5\degree\ for all directions of incoming light.
The field of view of a single telescope is 30\degree\ in azimuth and 28.6\degree\ in elevation.
The aperture stop with a an annular corrector lens controls the spherical aberration of the image maintaining an optical spot size of 0.5\degree. 
An optical filter, which matched to the nitrogen fluorescence light (approximately 300~nm to 400~nm) to reduce night-sky noise, also serves as a window over the aperture.
The PMTs are calibrated using a uniformly illuminated drum that covers the telescope aperture.
The illumination of the drum is compared to a standard photodiode from the US National Institute for Standards and Technology. 
Instruments for atmospheric monitoring include lasers, lidar systems, horizontal attenuation monitors, cloud monitors, star monitors and balloon-borne radiosondes.
The Central Laser Facility provides a steerable laser that can be seen by all fluorescence telescopes, which is the most useful for atmospheric monitoring.
At the present time three quarter of instruments have been completed and are in operation.
Detailed description of the Auger Observatory can be found elsewhere\cite{PAO}.

The greatest advantage of the Auger Observatory is its capability to study and understand the detailed performance of the instruments using hybrid events where the same shower is observed by both techniques.
In particular, the hybrid analysis benefits from the better energy measurement of the fluorescence technique and the uniformity of the surface detector aperture.
The angular resolution can be improved to a better accuracy than the surface array or the single fluorescence telescope could achieve independently.

Fig.~\ref{fig:PAO-hybrid} shows an example of a hybrid event with an energy of about $8\times10^{18}$~eV.
The top left panel shows the array hit distribution.
The solid line through the hit tanks represents the intersection of the plane formed by the shower and the fluorescence detector with the array.
The top right panel shows the lateral density distribution in the surface array for this event as a result of the fit to a lateral distribution function to determine the core position of the shower.
The particle density at the radius of 1000~m from the core, S(1000), can be used to estimate the energy of the shower.
The bottom left panel shows the energy measurement from the light collected by the fluorescence detector as a function of the slant depth.
This measurement determines the longitudinal development of the shower, whose integral is proportional to the total energy of the shower.
The bottom right panel shows the hit timing from both surface array and fluorescence detector as the shower sweeps out the angle $\chi$ formed by the shower and the fluorescence detector.
The axis of the air shower is determined by minimizing a $\chi^2$ function in this plot.
The resulting angular resolution is 0.6\degree.

\begin{figure}
\centering
\begin{tabular}{cc}
\parbox{6.5cm}{\centering
\includegraphics[height=6cm]{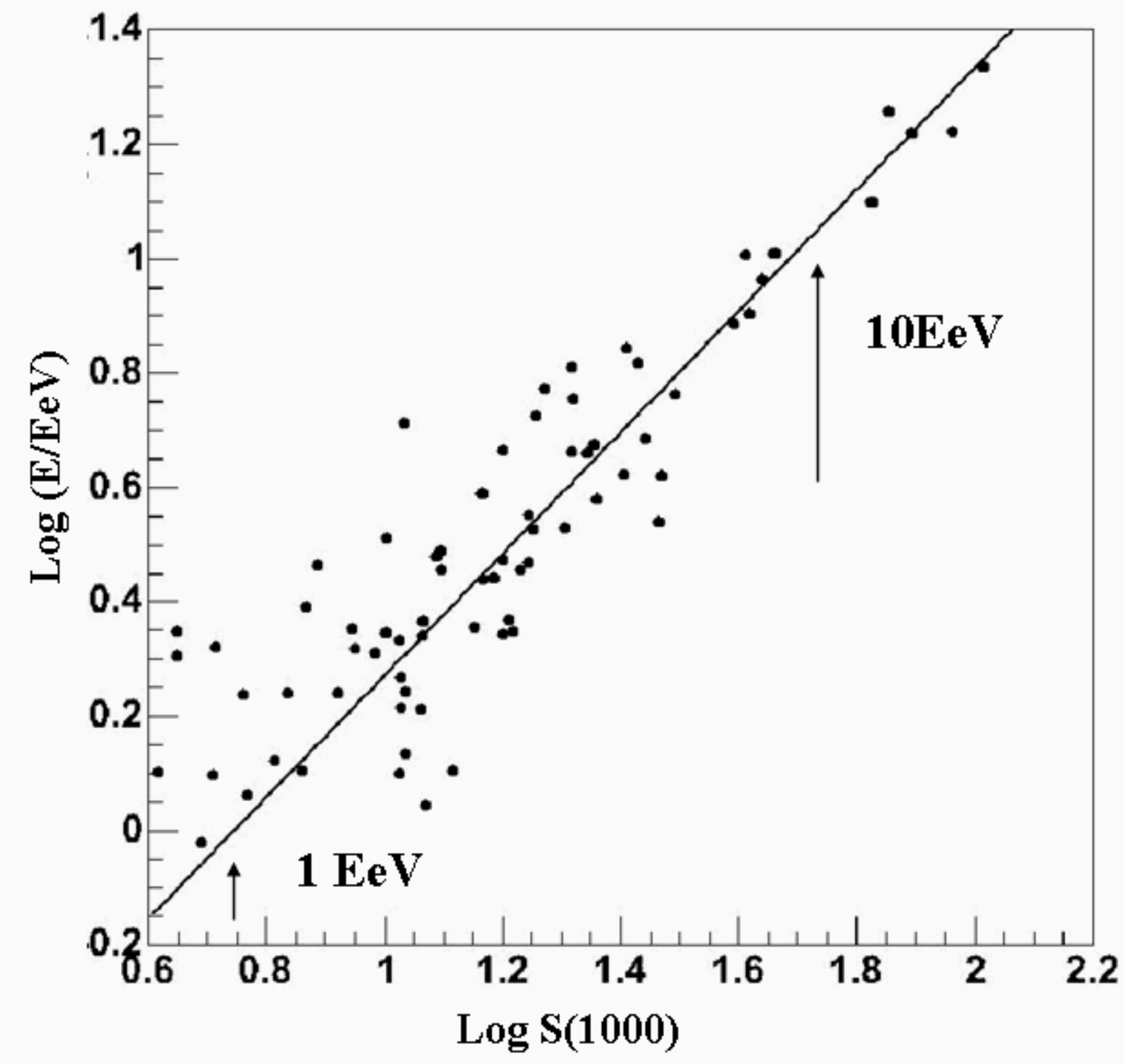}
\caption[The correlation of the log of the energy determined by the fluorescence technique and the log of S(1000) with a zenith angle correction.]
{\label{fig:PAO-S1000}The correlation of the log of the energy determined by the fluorescence technique and the log of S(1000) with a zenith angle correction.}
}
&
\parbox{9.7cm}{\centering
\includegraphics[height=6cm]{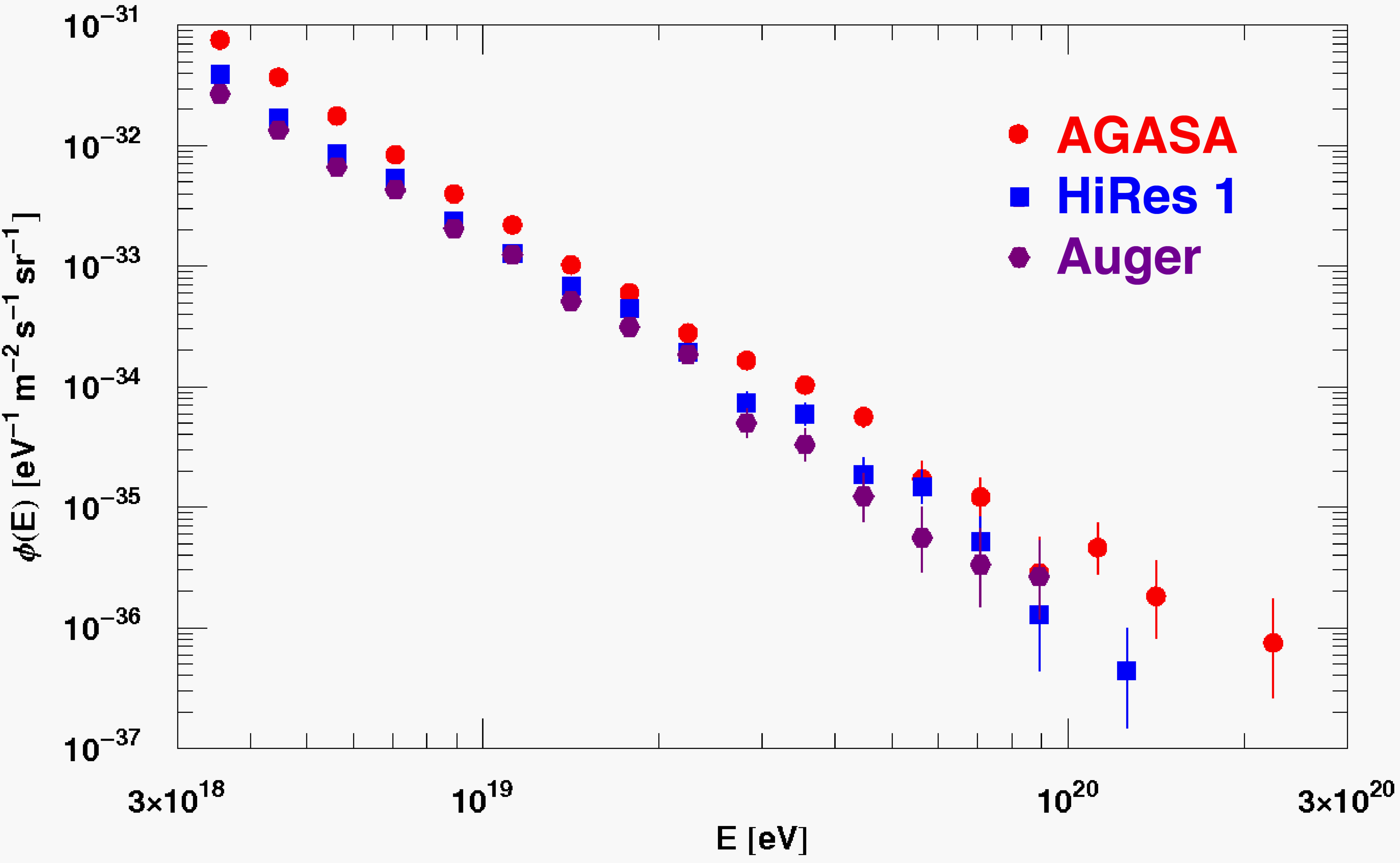}
\caption[Comparison of the Auger spectrum with those of AGASA and HiRes 1 (mono).]
{\label{fig:PAO-spectrum}Comparison of the Auger spectrum with those of AGASA and HiRes 1 (mono).}
}
\end{tabular}
\end{figure}

In order to calibrate the conversion factor from the S(1000) to the shower energy, a set of hybrid events was selected in which the fluorescence track length was at least 350 g$\cdot$cm${}^{-2}$ and \Cherenkov\ contamination was less than 10\%.
Fig.~\ref{fig:PAO-S1000} shows the correlation of the log of the energy determined by the fluorescence technique and the log of S(1000) with a zenith angle correction.
The energy conversion factor from S(1000) is obtained from the linear fit shown in the figure and can be applied to the surface-only events which accounts for majority of detected events due to the low duty cycle of the fluorescence telescopes. (They can be operational only during moonless nights.)

The first data set consists of $\sim$180,000 surface events and $\sim$18,000 hybrid events collected from January 1, 2004 through June 5, 2005. 
The acceptance was about 1750 \kmsq$\cdot$sr$\cdot$yr.
Fig.~\ref{fig:PAO-spectrum} shows the cosmic-ray energy spectrum observed by the Auger Observatory in comparison with those of AGASA and HiRes 1 (mono)\cite{PAO-result}.
It appears that the Auger result is consistent with that of HiRes and GZK suppression, although it is not conclusive.

\section{PAMELA}
PAMELA (Payload for Antimatter Matter Exploration and Light-nuclei Astrophysics) is a sattelite-borne instrument designed to study charged cosmic rays in the MeV--TeV region.
In particular, evidence of dark matter annihilation or stringent limit on the mass cross-section of dark matter can be derived from precise measurements of antiparticle (antiproton and positron) flux.
In addition, PAMELA will search for light antimatter such as anti-hellium, which is directly connected with the baryon-antibaryon asymmetry in the Universe.
The design goal of the PAMELA's performance is summarized in Table \ref{table:PAMELA-design}.
\begin{table}
\centering
\begin{tabular}{|c|c|c|}
\hline
Particle type & Number of events expected & Energy range \\ \hline
positrons & $3\times10^5$ & 50 MeV -- 270 GeV \\
antiprotons & $3\times10^4$ & 80 MeV -- 190 GeV \\
electrons & $6\times10^6$ & 50 MeV -- 2 TeV \\
protons & $3\times10^8$ & 80 MeV -- 700 GeV \\
light nuclei (up to Z=6) & & 100 MeV/$n$ -- 200 GeV/$n$ \\
light isotopes (${}^2$H, ${}^3$He) & & 100 MeV/$n$ -- 200 GeV/$n$ \\
limit on antinuclei & & $\overline{\mathrm He}/{\mathrm He} \sim 10^{-8}$ \\
\hline
\end{tabular}
\caption[Design goals of the PAMELA instrument.]
{\label{table:PAMELA-design}Design goals of the PAMELA instrument.}
\end{table}

\begin{figure}
\centering
\begin{tabular}{ccc}
\parbox{7cm}{\centering
\includegraphics[height=6cm]{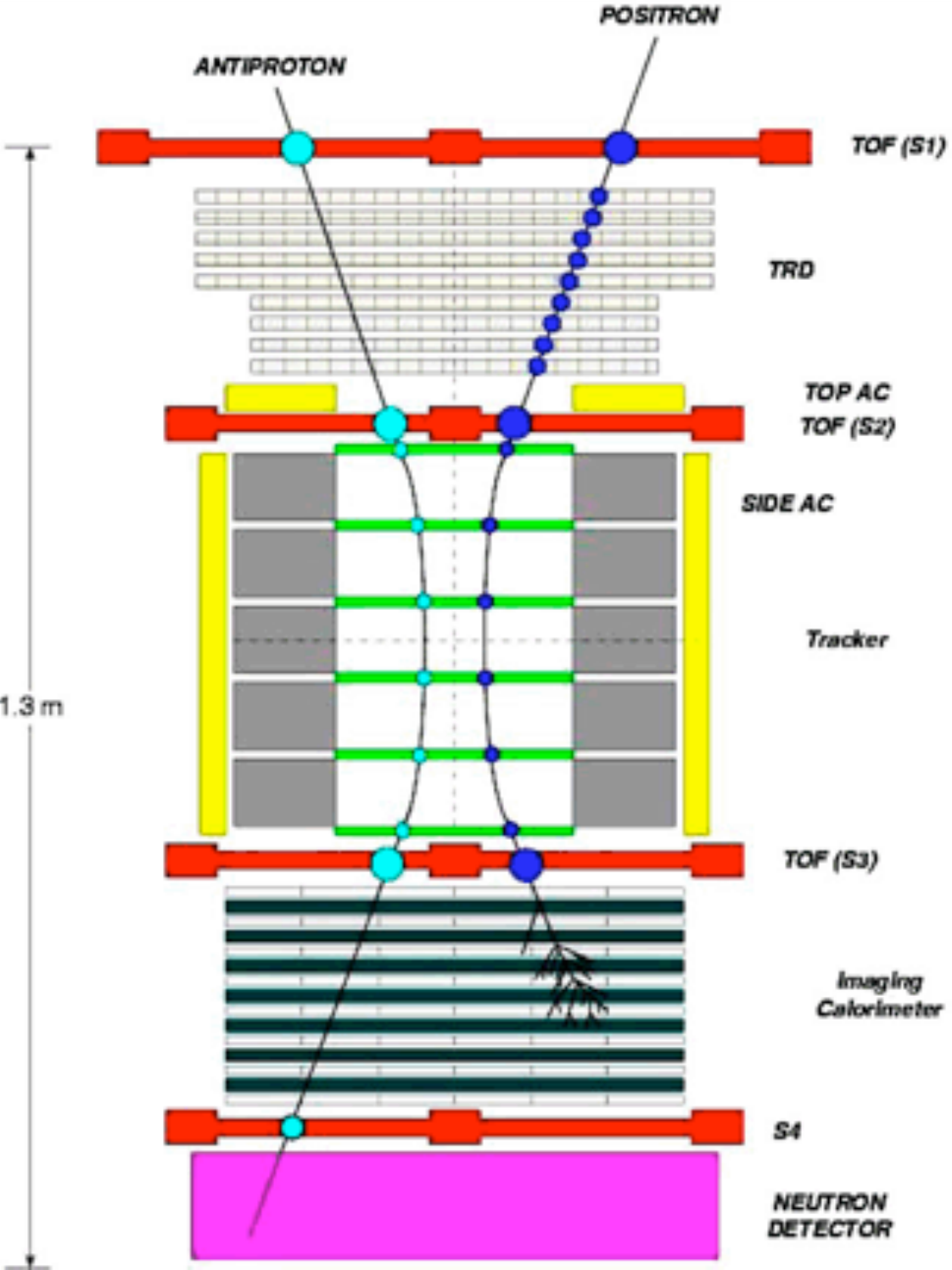}
\caption[A schematic view of the PAMELA instrument.]
{\label{fig:PAMELA-instrument}A schematic view of the PAMELA instrument.}
}
& & \parbox{9cm}{\centering
\includegraphics[height=6cm]{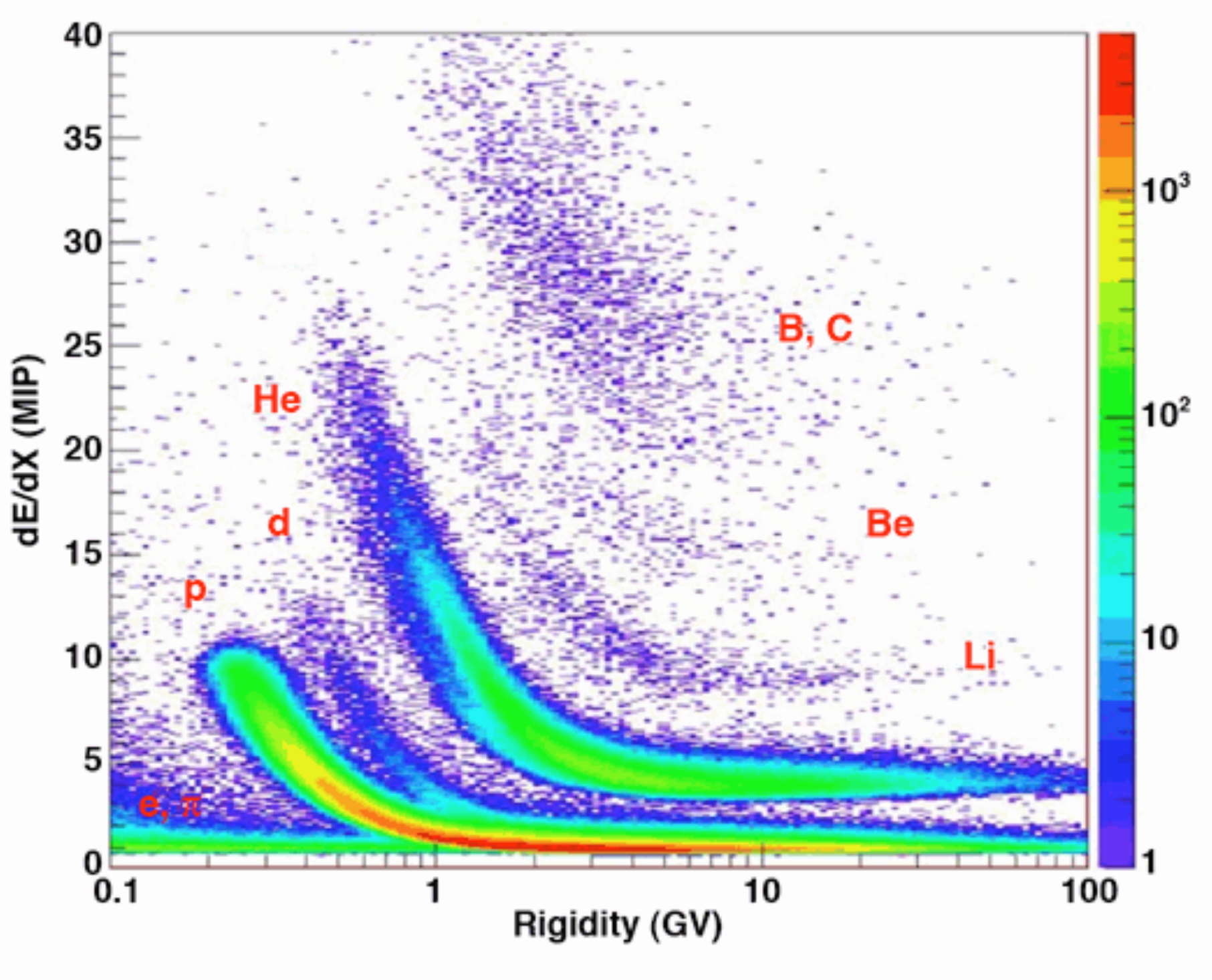}
\caption[Correlations of the energy loss in the DSSDs and the rigidity measured by the spectrometer.]
{\label{fig:PAMELA-dEdX}Correlations of the energy loss in the DSSDs and the rigidity measured by the spectrometer.}
}
\end{tabular}
\end{figure}

The PAMELA instrument has dimensions of approximately $75\times75\times123$~cm${}^3$, an overall mass of 450~kg and a power consumption of 355~W. 
The overall acceptance is 21.5 \cmsq$\cdot$sr defined by the spectrometer geometry.
These factors are constrained by the capability of the launch vehicle (Soyuz-U rocket) and the spacecraft (Resurs DK1 earth-observation satellite).
The PAMELA instrument consists of a time-of-flight (TOF) system, a magnetic spectrometer, an anticoincidence system, an electromagnetic imaging calorimeter, a shower tail catcher scintillator and a neutron detector as shown in Fig.~\ref{fig:PAMELA-instrument}.

Plastic scintillators are placed above and below the spectrometer to form a TOF system.
The timing resolution of the TOF system is sufficient to eliminate the upward-going particles and separate protons and electrons below $\sim$1~GeV/$c$.
The TOF also provides ionizing energy loss measurements for absolute charge determination and the primary trigger for the instrument.

The spectrometer consists of a 0.43~T permanent dipole magnet and 6 layers of double-sided silicon strip detectors (DSSD) and measures the momentum and the sign of the charge.
In addition, the ionizing energy loss measurements in the DSSDs provides the absolute value of the particle charge.
Fig.~\ref{fig:PAMELA-dEdX} shows correlations of the energy loss in the DSSDs and the rigidity measured by the spectrometer in Space.
Hydrogen, deuterium and helium can be clearly identified with some indication of lithium, beryllium, boron and carbon.
The magnet is enclosed by ferromagnetic shielding to minimize the stray magnetic field that could interfere with the satellite instruments and navigation systems.
Each layer is built from three DSSD ladders that comprises two sensors with a size of $5.33\times7.00$~\cmsq.
The strip pitch of the DSSD is 25~\micron\ in the X view (bending) and 67~\micron\ in the Y view (non-bending).
The signal of the Y view is extracted via the second metal traces orthogonal to the implant.
The strips are biased via punch-through on the junction side (X view) and polysilicon resistors ($>10$~M$\Omega$) on the ohmic side (Y view).
The fine position resolution (3--7.5~\micron\ in X and 8--13~\micron\ in Y) afforded by the DSDDs corresponds to the maximum detectable rigidity of $\sim$1.2~TeV/$c$.
Due to the presence of the resolution tail and the large particle background (the antiparticle/partilce ratio is of the order of $10^{-4}$), the maximum momentum to identify the charge is about 190~GeV/$c$ for antiprotons and 270~GeV/$c$ for positrons.
Each ladder is connected with a front-end hybrid circuits that features VA1 ASIC (Application Specific Integrated Circuits).
The VA1 contain 128 channels of charge sensitive preamplifiers, CR-RC shapers and sample/hold circuits followed by an analog multiplexer.
The VA1 is operated with less than the nominal drain current yielding a power consumption of 1~mW/channel to meet the power constraint.
The ENC (equivalent noise charge) is measured in Space to be 510 electrons on the junction side and 1090 electrons on the ohmic side, which is almost identical to the values obtained on the ground.

The electromagnetic calorimeter consists of 44 layers of single-sided silicon detectors interleaved with 22 plates of tungsten absorber with 0.74 radiation lengths, giving a total depth of 16.3 radiation lengths and $\sim$0.6 nuclear interaction lengths. 
The $8\times8$~\cmsq\ silicon detectors are segmented into 32 readout strips with a pitch of 2.4 mm.
The silicon detectors are arranged in a $3\times3$ matrix and each of the 32 strips is daisy-chianed to the corresponding strip on the other two detectors in the same row (or column), thereby forming 24 cm long readout strips.
The orientation of the strips of two consecutive layers is orthogonal to provide two-dimensional spatial information.
The three-dimensional measurement of the shower shape, combined with the measurement of the particle energy loss in each silicon detector, yields a high identification power for electromagnetic showers. 
The calorimeter is found to have a proton rejection factor of about $10^5$ while keeping about 90\% electron (and positron) efficiency. 
The electron (and positron) energy measurement by the calorimeter will facilitate a cross-calibration with the magnetic spectrometer.

As described above, PAMELA employs conventional technologies for silicon sensors and readout electronics, which is important to avoid any delays associated with relatively new technologies.
Delays in satellite-borne experiments is very costly.
Conventional technologies also tend to increase the reliabilities of components, which is crucial in space experiments for obvious reasons.
A detailed description of the PAMELA instrument can be found in ref \cite{PAMELA}.

\section{IceCube}
\begin{figure}[bh]
\centering
\begin{tabular}{ccc}
\parbox{8cm}{\centering
\includegraphics[height=6cm]{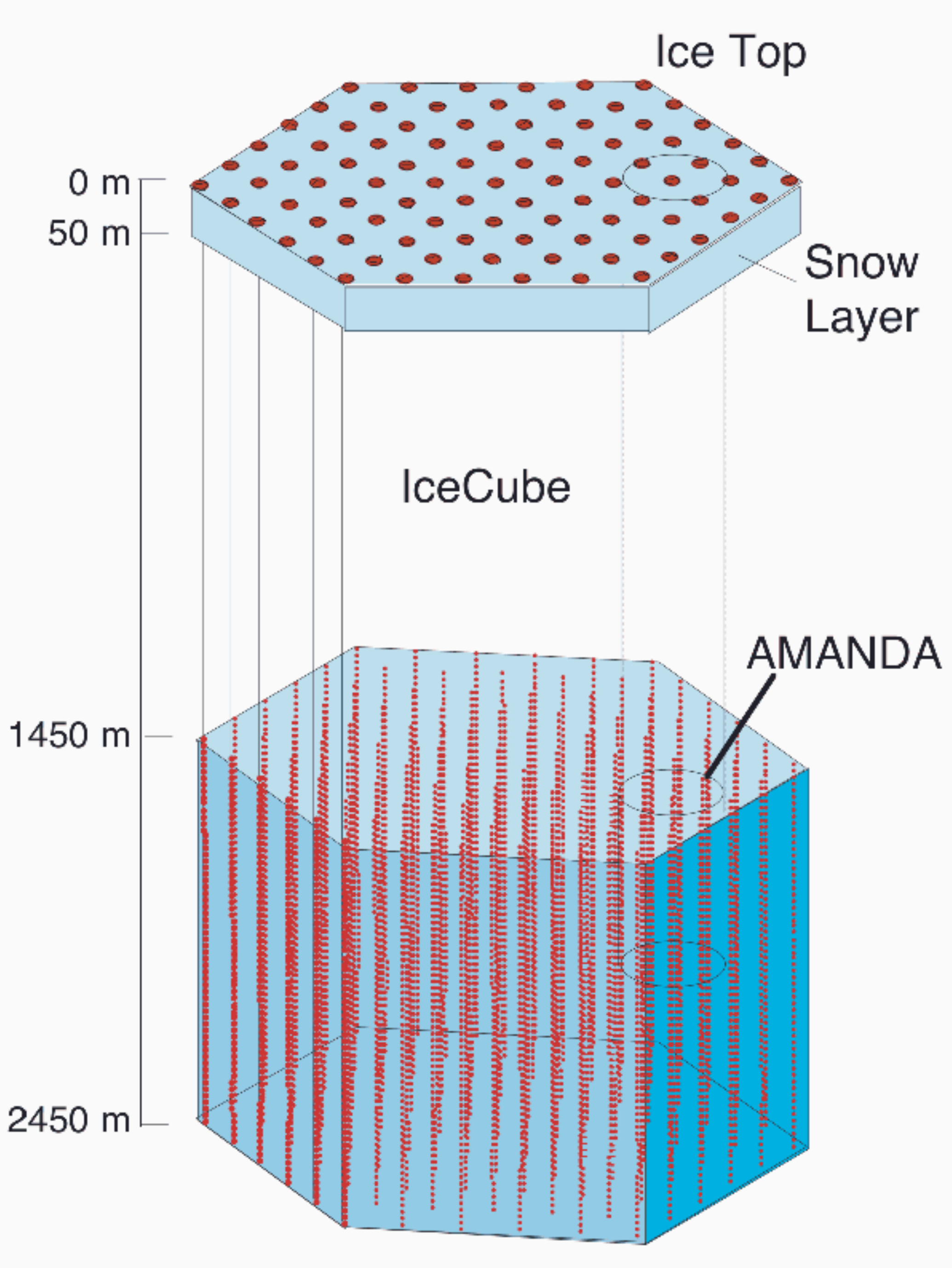}
\caption[A schematic view of the IceCube Observatory.]
{\label{fig:IceCube-schematic}A schematic view of the IceCube Observatory.}
}
& &
\parbox{8cm}{\centering
\includegraphics[height=6cm]{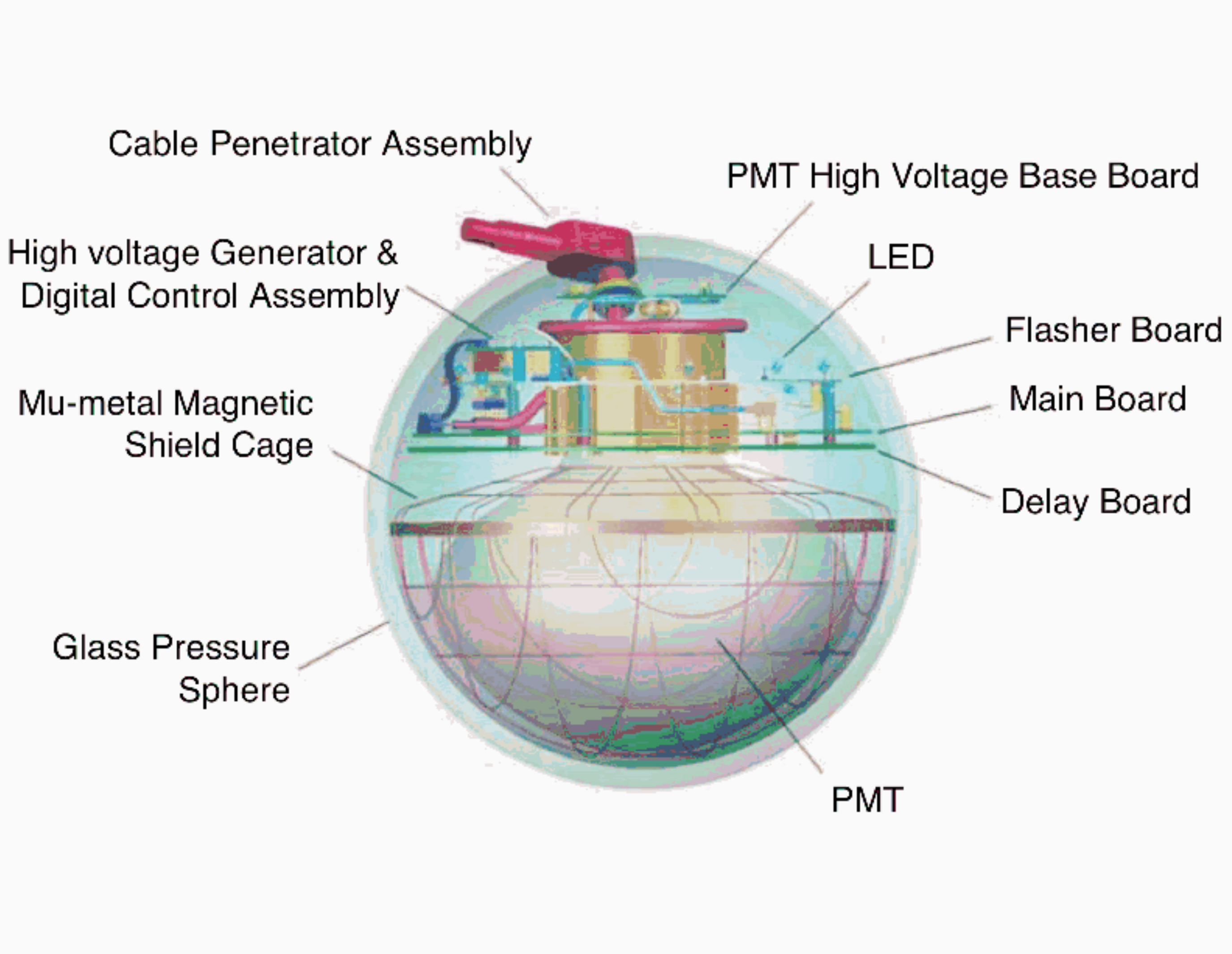}
\caption[A schematic view of a Digital Optical Module.]
{\label{fig:IceCube-DOM}A schematic view of a Digital Optical Module.}
}
\end{tabular}
\end{figure}

The IceCube neutrino observatory is a next-generation cubic-kilometer scale high-energy cosmic neutrino telescope currently under construction at the South Pole.
The construction started in January 2005 and part of the detectors are in operation since February 2006.
IceCube is designed to detect astronomical neutrinos in an energy range from a few $10^{11}$~eV to $10^{17}$~eV.
\Cherenkov\ photons emitted from relativistic charged particles in 3~km thick glacial ice are detected by an array of Digital Optical Modules (DOMs).
IceCube consists of 80 strings with 60 DOMs in 17~m spacing between a depth of 1450 and 2450~m where the glacial ice is transparent as illustrated in Fig.~\ref{fig:IceCube-schematic}.
The strings are arranged in a hexagonal lattice pattern with a spacing of approximately 125~m.
The deep-ice DOMs are deployed along electrical cable bundles which comprises 30 twisted pair copper cables packaged in 15 twisted quads and carry power and information between the DOMs and surface electronics.
DOMs are also placed at the surface near the top of each string which forms an air shower array called IceTop to study the atmospheric muon background reliably.

Each DOM encloses a 10" R7081-02 PMT made by Hamamatsu Photonics in a transparent glass pressure vessel along with a high voltage system, a LED flasher board for optical calibration in ice and a digital readout board, making it an autonomous data collection unit as shown in Fig.~\ref{fig:IceCube-DOM}.
All electronics of the DOMs were designed for high reliability and stress testing was performed to screen for high quality.
Low power consumption was a design goal, the power of a single DOM being 3W in normal operation.

The PMT signal is amplified at three different gains ($\times1/4$, $\times2$ and $\times16$) to obtain a sufficient dynamic range.
The signals are digitized by a fast analog transient waveform recorder at a sampling rate of 300~MHz.
The linear dynamic range of the sensor is 400 photoelectrons in 15~ns and the integrated dynamic range is of more than 5000 photoelectrons in 2~ms.
The digitization is performed only when it has a local trigger with a neighboring DOM.
The waveforms along with GPS time stamps are sent from each DOM to the surface data acquisition system, where data are time sorted.
The digital transmission of PMT signals in IceCube is the most critical improvement from the predecessor, AMANDA, where the analog signals from PMTs are propagated through the cables to the surface data acquisition, resulting in difficulties in the calibration.
The digital transmission also realizes the higher dynamic range.
Detailed description of the IceCube Observatory can be found in ref \cite{IceCube}.

A Monte Carlo simulation of a realistic model of the IceCube was performed to study the effective area and the angular resolution and to estimate the sensitivity to predicted fluxes of muon neutrinos at TeV to PeV energies from AGNs (Active Galactic Nuclei) and GRBs (Gamma-Ray Bursts) \cite{Ahrens}.
About 3,300 triggers per year is expected from muons induced by astrophysical neutrinos while 0.8 million triggers per year are expected from atmospheric neutrinos.
The event selection criteria are designed to reduce the misreconstructed downward-going muon tracks to the level of atmospheric neutrinos while retaining roughly 1/3 of the signal events.
The resulting effective area is found to be greater than 1 \kmsq\ above $10^{13}$~eV and go up to 1.4 \kmsq\ at $10^{17}$~eV.
The effective area is almost uniform up to the horizon (upward-going) and deteriorates drastically above the horizon (downward-going) at energies below $10^{14}$~eV.
At energies above $10^{15}$~eV, the degradation of the effective area is more mild and about one half of the peak at the zenith, which means that IceCube can observe a large part of the Galaxy, including the Galactic center.
The angular resolution is found to be around 0.8\degree\ for energies above $10^{13}$~eV.
\begin{figure}[htb]
\centering
\includegraphics[height=6cm]{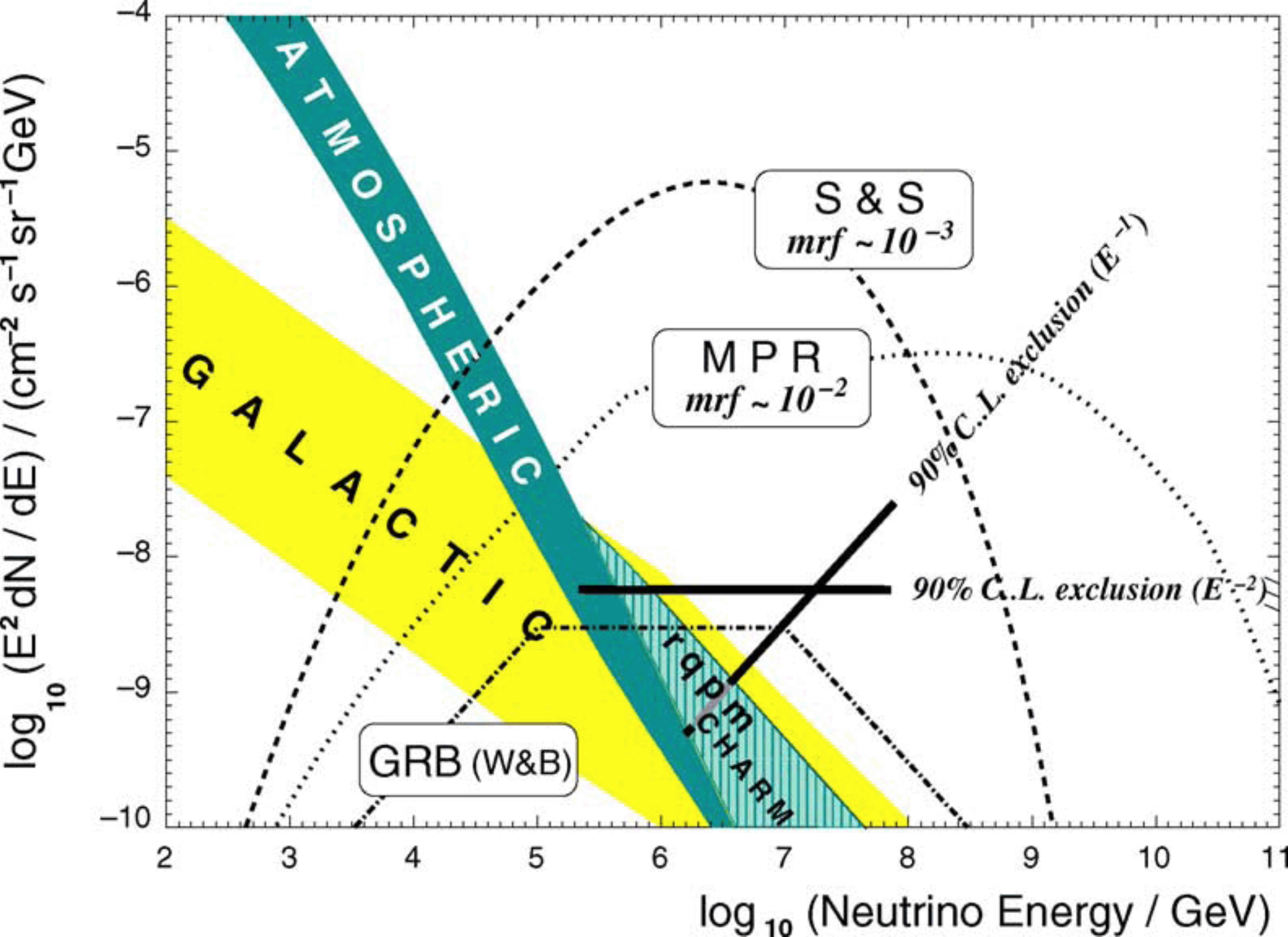}
\caption[Expected sensitivity of the IceCube detector to neutrino flux from point sources. Expected neutrino flux from some AGN and GRB models are also shown.]
{\label{fig:IceCube-sensitivity}Expected sensitivity of the IceCube detector to neutrino flux from point sources. Expected neutrino flux from some AGN and GRB models are also shown.}
\end{figure}

Given the above estimates for background rates, effective area and angular resolution, the 5$\sigma$ sensitivity is calculated to be $E^2\cdot dN/dE = 10^{-8}$~\flux\ for diffuse sources with $E^{-2}$ spectra and $7\times 10^{-9}$~\flux\ for point sources after three years of observation, which is two orders of magnitude below present experimental limits.
Fig.~\ref{fig:IceCube-sensitivity} shows the comparison of the expected sensitivity and model predictions for AGN\cite{MPR},\cite{SS} and GRB\cite{WB}.
It indicates that IceCube can probe the validity of these AGN models in less than three years.
In the case of GRBs, the sensitivity is much higher than indicated due to the short duration ($<100$~s) of the GRBs (number of background events are reduced by a factor of $\sim$80,000).
IceCube is expected to observe at least one 5$\sigma$ signal event out of 200 GRBs.

\section{ANITA}
\begin{figure}[bh]
\centering
\begin{tabular}{ccc}
\parbox{8cm}{\centering
\includegraphics[height=6cm]{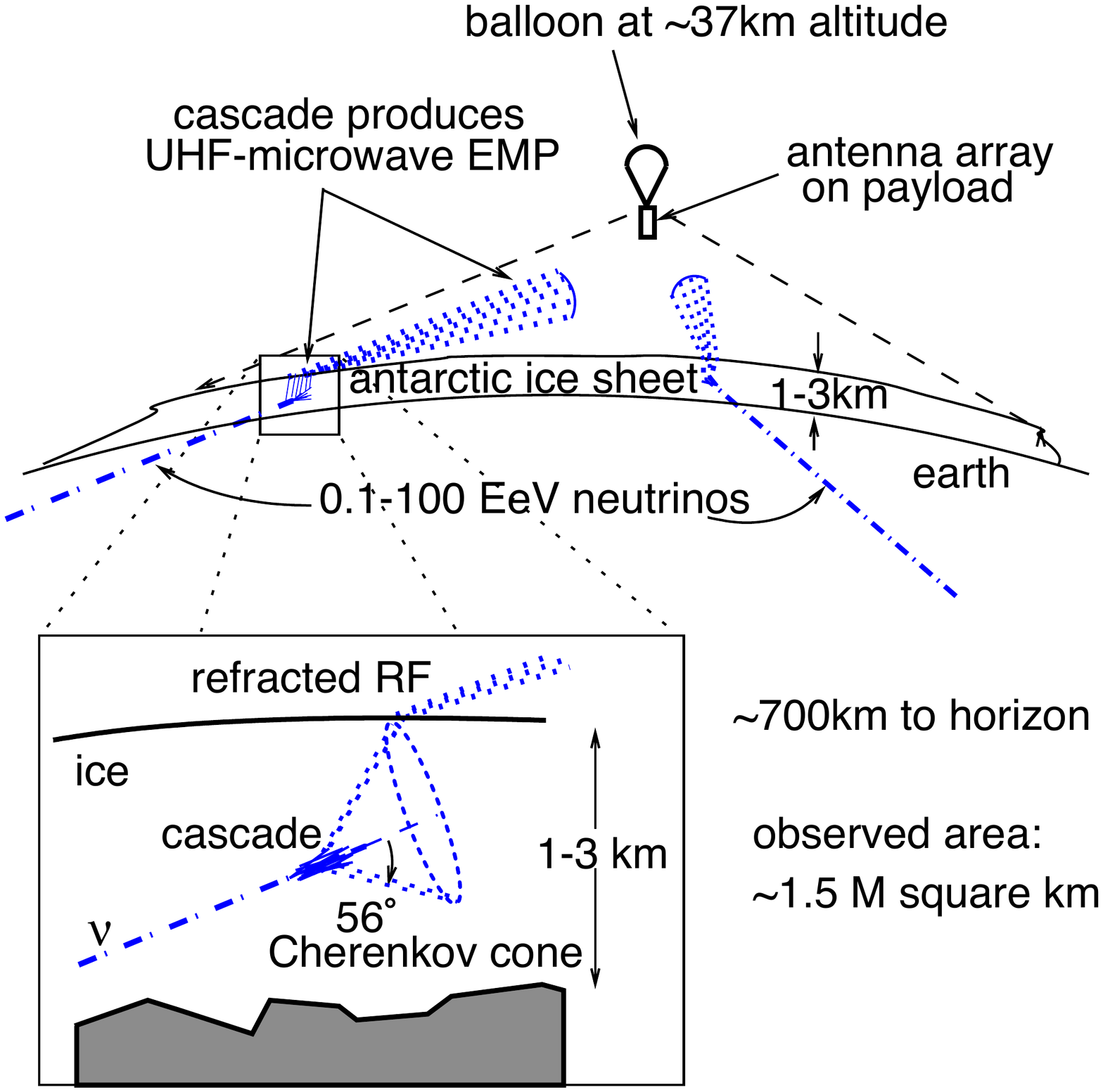}
\caption[Principle of the ANITA experiment.]
{\label{fig:ANITA-concept}Principle of ANITA experiment.}
}
& &
\parbox{8cm}{\centering
\includegraphics[height=6cm]{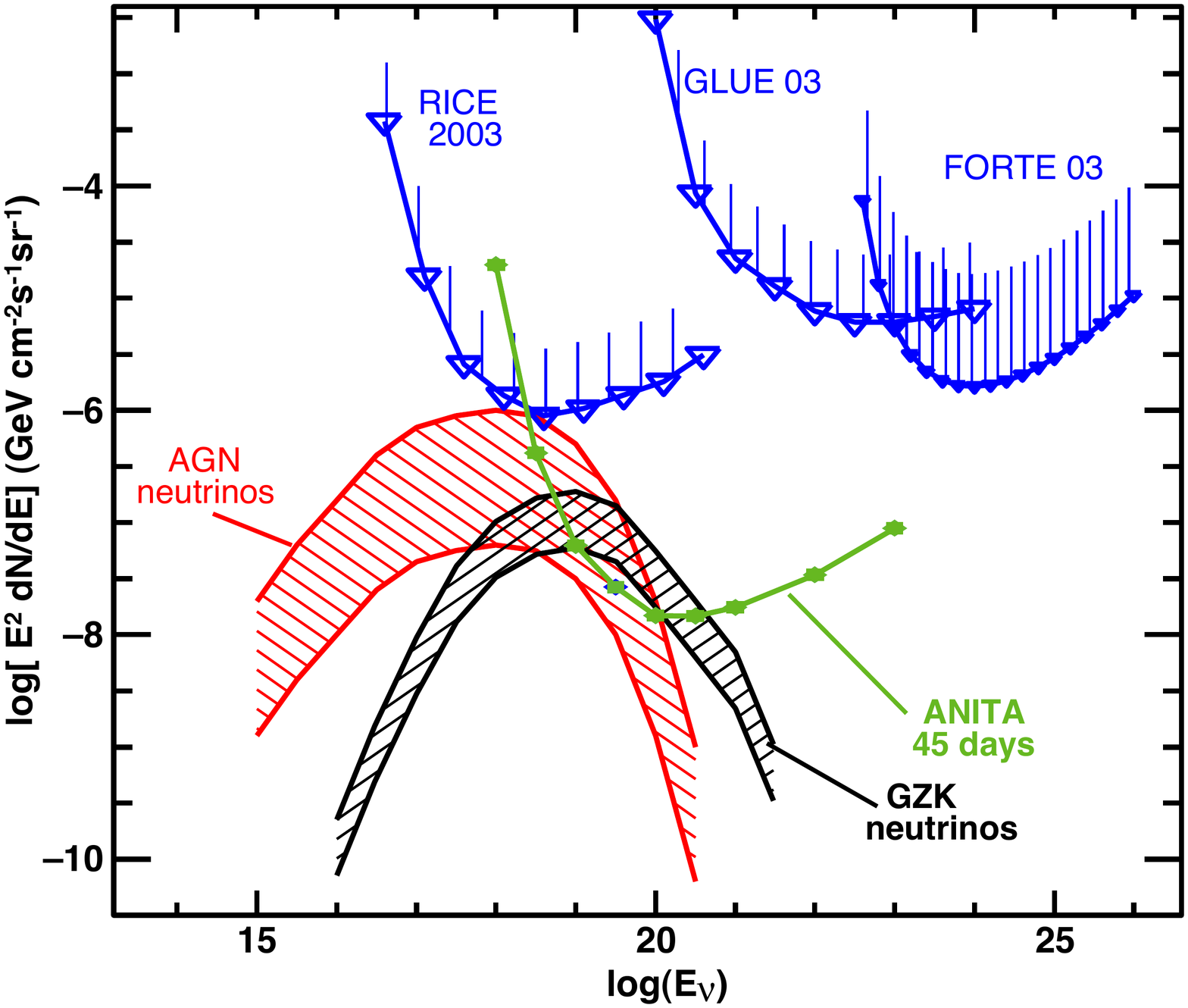}
\caption[Expected sensitivity of the ANITA experiment compared with the predicted flux of GZK neutrinos.]
{\label{fig:ANITA-sensitivity}Expected sensitivity of the ANITA experiment compared with the predicted flux of GZK neutrinos.}
}
\end{tabular}
\end{figure}

The ANITA (Antarctic Impulsive Transient Antenna) experiment is designed to detect UHE neutrinos from GZK interactions of UHECRs with CMB photons, which requires an enormous target volume, of the order of $10^{3}$~km${}^3\cdot$sr.
The ANITA takes advantage of the coherent impulse radio \Cherenkov\ emission from the charge asymmetry ($\sim$20\%) in an electromagnetic shower induced by a UHE neutrino as predicted by Askaryan\cite{Askaryan}.
In dense media, the shower appears as a single charge moving through the dielectric at wavelengths of the shower size ($\sim$10~cm) or greater.
Since the radiated power for coherent \Cherenkov\ emission grows quadratically with the charge of the shower, its power in the cm-to-m wavelength is $\sim10^{13}$ times greater for a $10^{20}$~eV neutrino shower than the power from the incoherent emission in other wavelengths.
The balloon-borne radio telescope at an altitude of 35--40~km above Antarctica can detect such a radio emission at a distance allowing to use a large volume ($\sim$$10^6$~km${}^3$) of the radio-transparent ice as a target as illustrated in Fig.~\ref{fig:ANITA-concept}.
The shower angle with respect to the radio direction can be determined by the polarization of the radio pulse.
Fig.~\ref{fig:ANITA-sensitivity} shows the expected sensitivity of ANITA for 45~days of balloon flight.
The predicted neutrino flux from the GZK process of UHECRs is also shown\cite{ANITA}.
This technique has a great future potential since it can exploit huge volumes of natural targets using a relatively small detector.

\begin{figure}
\centering
\begin{tabular}{ccc}
\parbox{7.8cm}{\centering
\includegraphics[height=6cm]{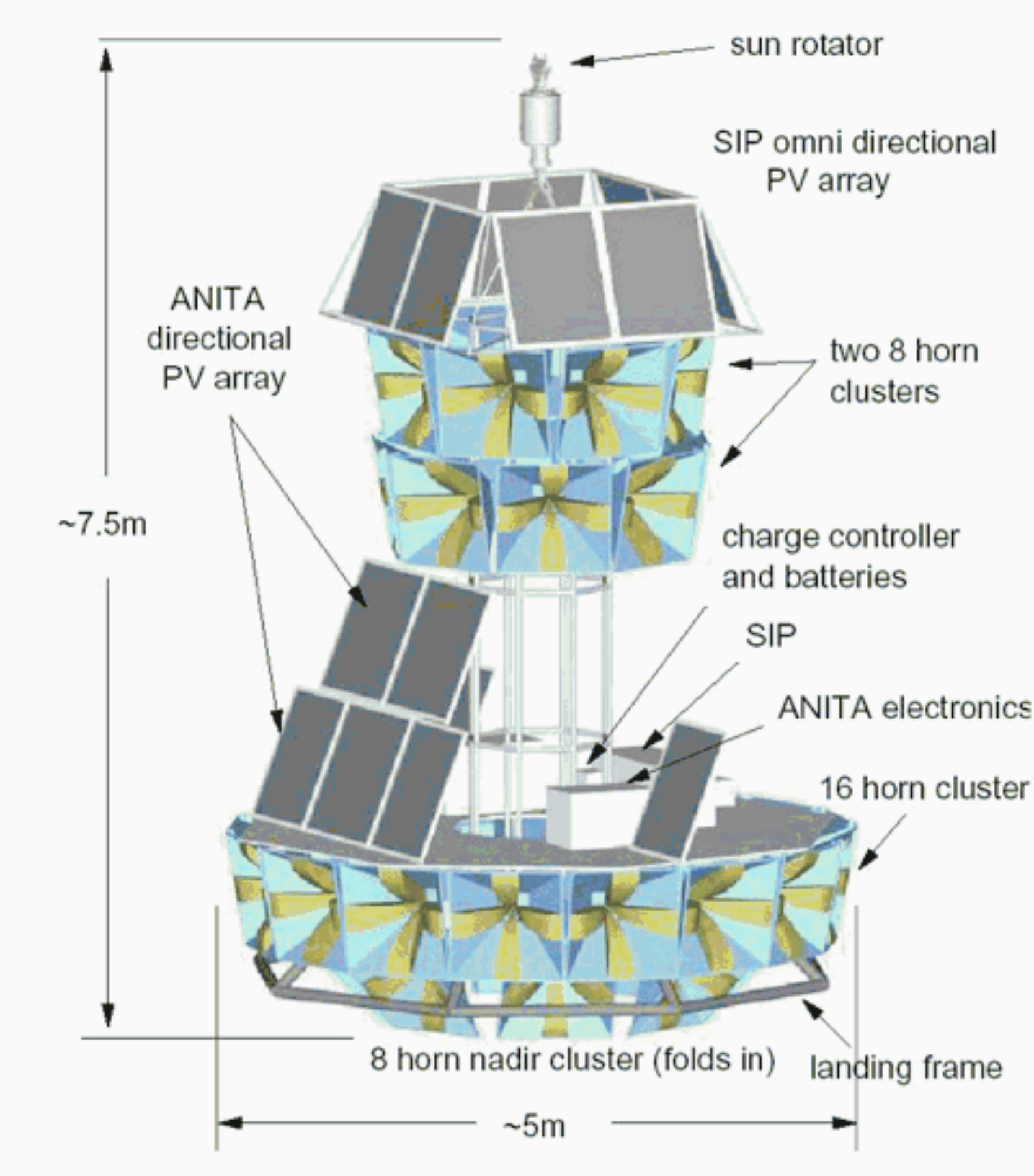}
\caption[A conceptual drawing of the ANITA experiment.]
{\label{fig:ANITA-payload}A conceptual drawing of the ANITA experiment.}
}
& &
\parbox{8.2cm}{\centering
\includegraphics[height=6cm]{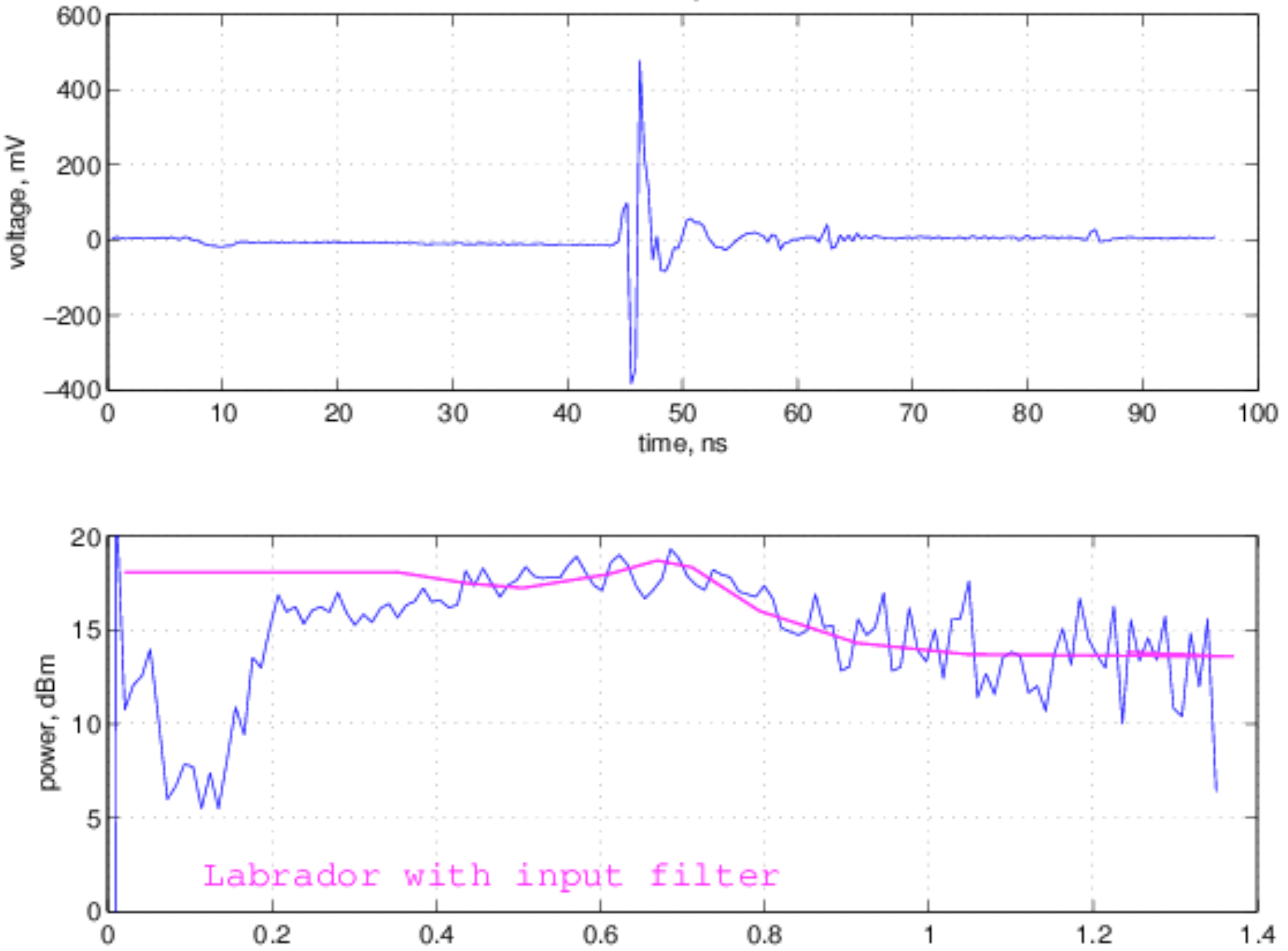}
\caption[Waveform sampling (top) and Fast Fourier
Transform (bottom) of the LAB3 ASIC response.]
{\label{fig:LAB3-FFT}Waveform sampling (top) and Fast Fourier
Transform (bottom) of the LAB3 ASIC response.}
}
\end{tabular}
\end{figure}

The ANITA payload consists of 32 dual-polarization, quad-ridged horn antennas arranged in cylindrically symmetric upper and lower clusters as shown in Fig.~\ref{fig:ANITA-payload}.
Each antenna records two linear polarizations of an incoming radio pulse and has a beam width of about 60\degree.
Quad-ridge horn antennas were chosen because of their excellent frequency response over the frequency band of interest, from 0.2--1.2~GHz (limited by the RF transmissivity of ice), and a very small phase dispersion, resulting in a sub-nanosecond impulse response.
The separation between upper and lower antenna clusters is required to determine the elevation angle of the radio pulse from the arrival time differences.
Eight additional vertically-polarized omnidirectional broadband monitor antennas (4 bicones and 4 discones) are used to complement the highly directive horns by providing pulse-phase interferometry response.
The radio pulse from the antenna is split into a trigger circuit and a waveform recorder.
In the trigger circuit, the full 1~GHz bandwidth is split into 4 separate frequency bands to minimize the EMI susceptibility.

The waveform recorder is a major challenge since it is required to operate at a rate of $\sim$3 GSa/s to record $\sim$1.2~GHz pulses with a tight power constraint.
Typical flash ADCs do not have sufficient dynamic range (8 bit) and consume too much power (a few Watts/channel).
A custom ASIC, LABRADOR (Large Analog Bandwidth Recorder And Digitizer with Ordered Readout)\cite{LAB3}, was designed to meet the specifications.
It is based on switched capacitor array circuits to record analog values and Wilkinson-type ADCs in each sampling capacitor to digitize the sampled waveform.
Fig.~\ref{fig:LAB3-FFT} demonstrates the performance of this device, where the top panel shows a raw band-limited waveform, and the bottom panel shows the Fourier transform, corrected for the measured spectral content of the test impulse.
The dip below 200~MHz is due to the high-pass filter upstream of the ASIC.

\begin{figure}
\centering
\begin{tabular}{ccc}
\parbox{8cm}{\centering
\includegraphics[height=6cm]{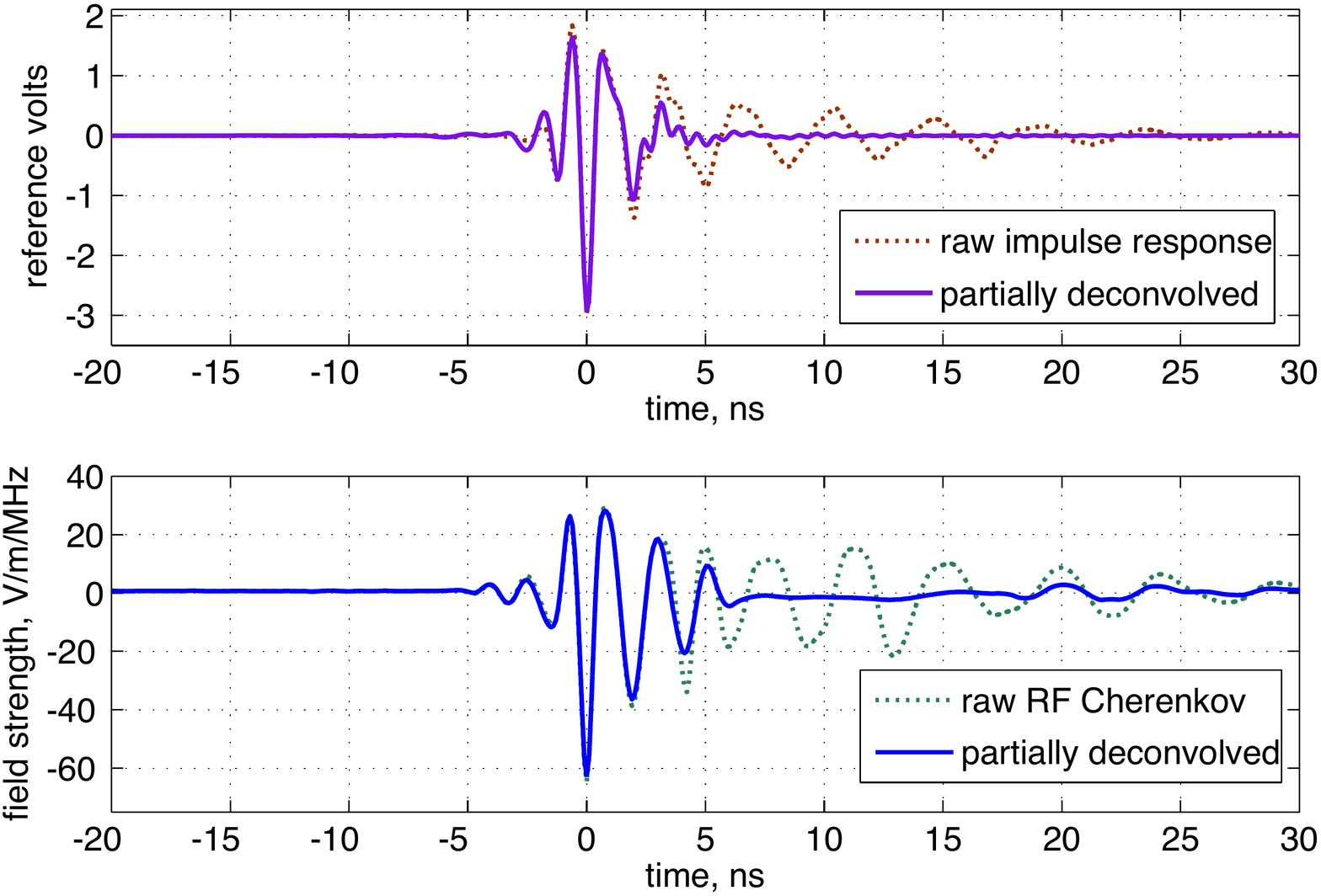}
\caption[(Top panel) Raw and partially deconvolved impulse response of the ANITa receiver system. (Bottom panel) Observed pulse in the SLAC T486 beam test.]
{\label{fig:ANITA-pulse}(Top panel) Raw and partially deconvolved impulse response of the ANITa receiver system. (Bottom panel) Observed pulse in the SLAC T486 beam test.}
}
& &
\parbox{8cm}{\centering
\includegraphics[height=6cm]{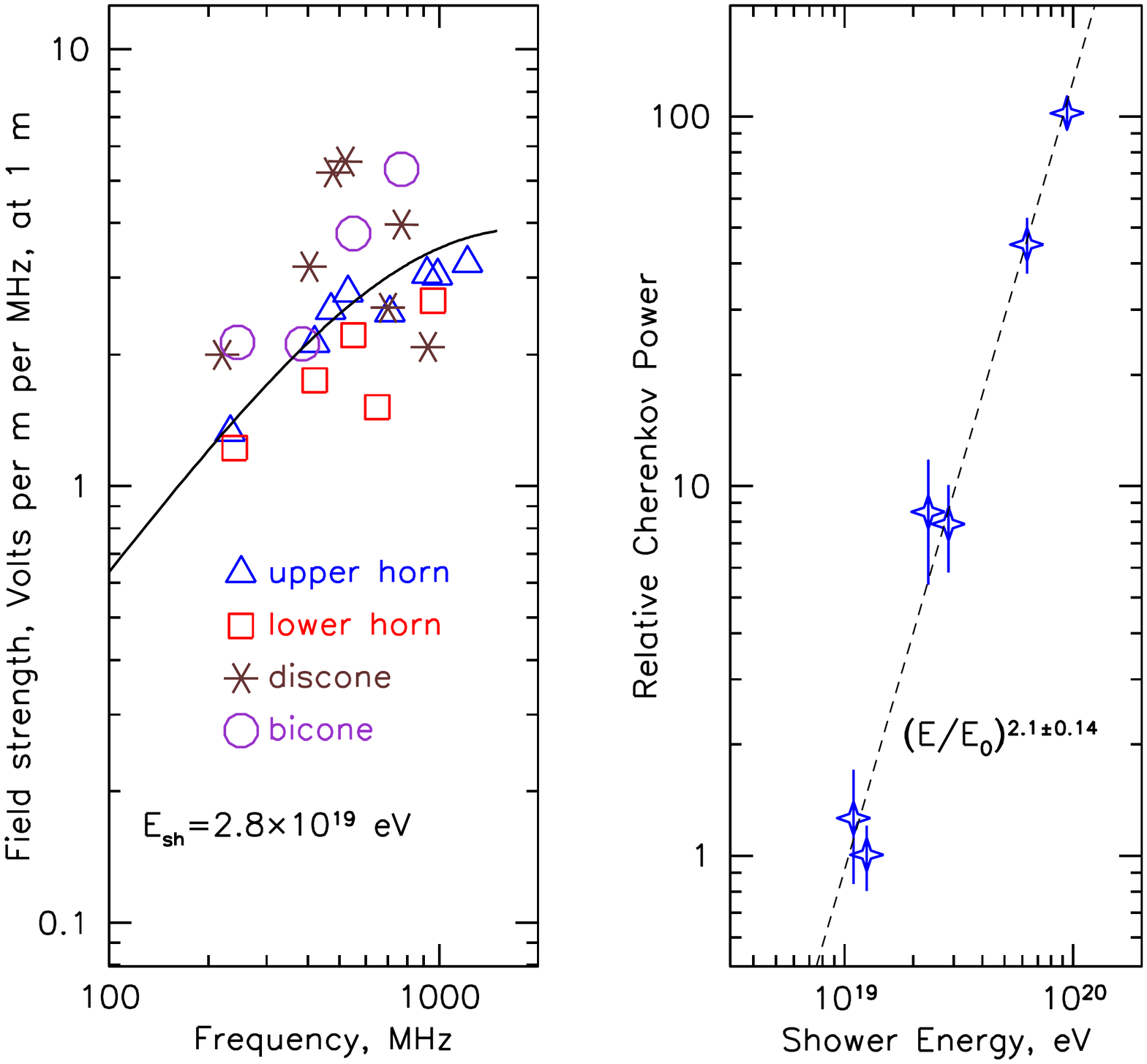}
\caption[(Left panel) Field strengh as a function of the radio \Cherenkov\ emission frequency in the T486 experiment. (Right panel) Quadratic dependence of the pulse power of the radiation detected in T486.]
{\label{fig:ANITA-beam-test}(Left panel) Field strengh as a function of the radio \Cherenkov\ emission frequency in the T486 experiment. (Right panel) Quadratic dependence of the pulse power of the radiation detected in T486.}
}
\end{tabular}
\end{figure}

In order to confirm the Askaryan effect in the ice, a beam test, SLAC T486, was carried out using the showers produced by 28.5~GeV electron beam with 10~ps bunches of $10^9$ electrons, corresponding to a composite energy of $3\times10^{19}$~eV.
This test is critical to the ANITA experiment since there is no natural background to calibrate the instrument due to its low backgrounds.
Since the primary beam is electrons, the charge asymmetry of the shower is expected to be $\sim$20\% which is 15\% higher than that from the neutral primary.
The top panel of Fig.~\ref{fig:ANITA-pulse} shows an example of the impulse response of the ANITA receiver system.
The bottom panel shows the waveform observed near the peak of the \Cherenkov\ cone in the beam test.
The apparent ringing in the both raw impulses is due to the group delay variation of the edge response of the bandpass filters.
the left panel of Fig.~\ref{fig:ANITA-beam-test} shows the absolute field strength measured by several different antennas as a function of the radio \Cherenkov\ emission frequency.
The uncertainty in these data are dominated by uncertainties in the gain calibration of the antennas and in removing secondary reflections from the measured impulse power.
The curve in the figure shows the result of a parameterization based on shower+electrodynamics simulations for ice, which is in a good agreement with the experimental results within errors.
The right panel of Fig.~\ref{fig:ANITA-beam-test} shows the measured \Cherenkov\ power as a function of the shower energy, which clearly demonstrates a quadratic dependence.
a detailed description of the beam test can be found in ref \cite{ANITA-beam-test}.

The first Antarctica balloon flight of ANITA is carried out from December 15, 2006 through January 19, 2007 (total of $\sim$35 days).
The payload was successfully recovered.
We expect that the result of this flight will yield the observation of several GZK neutrinos or in severe limits on modeling of the GZK process.

\section{Imaging Atmospheric \Cherenkov\ Telescopes}
\begin{figure}[bh]
\centering
\begin{tabular}{ccc}
\parbox{8cm}{\centering
\includegraphics[height=6cm]{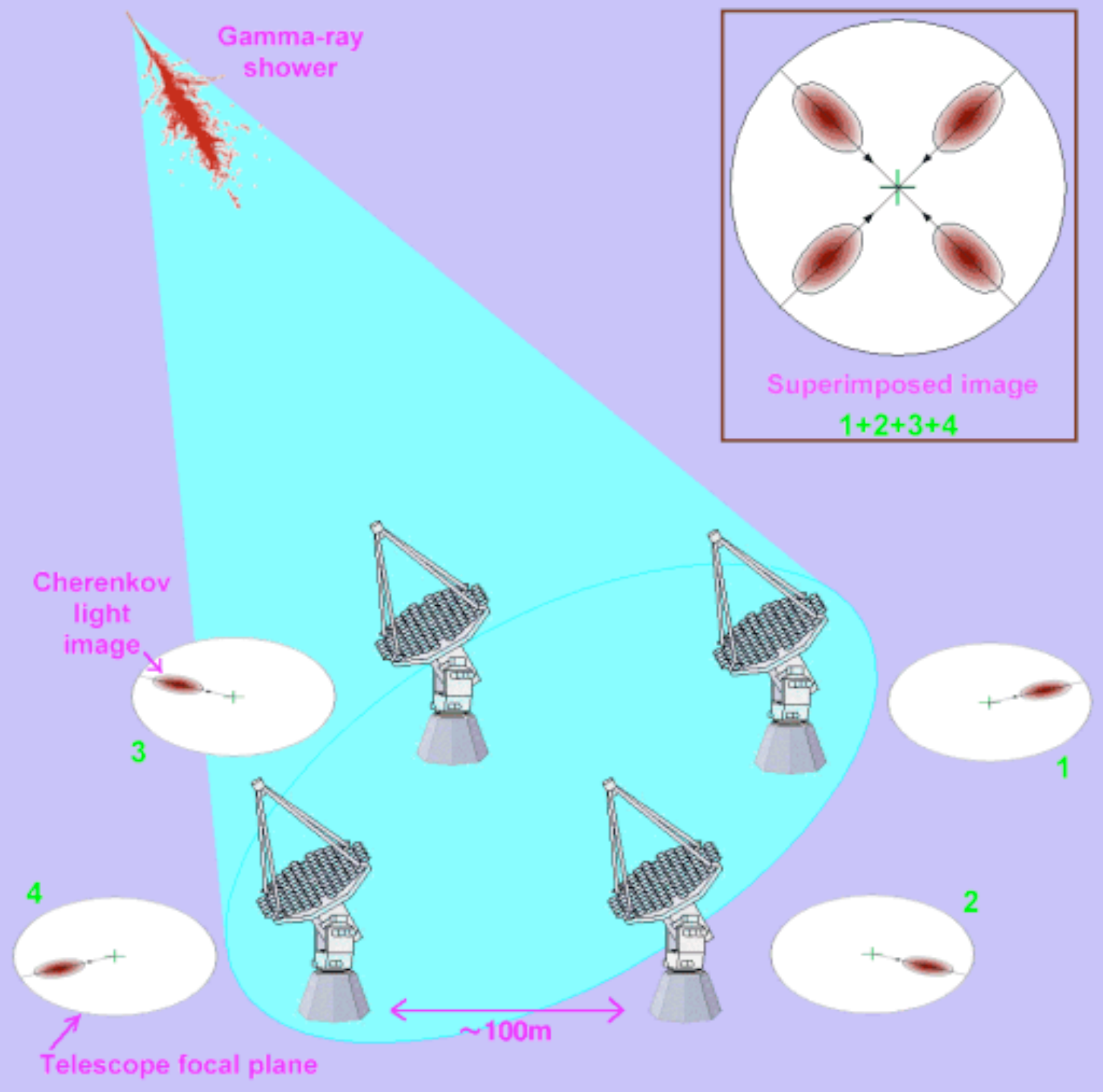}
\caption[Conceptual view of multi-telescope operation of IACT.]
{\label{fig:IACT-stereo}Conceptual view of multi-telescope operation of IACT.}
}
& &
\parbox{8cm}{\centering
\includegraphics[height=6cm]{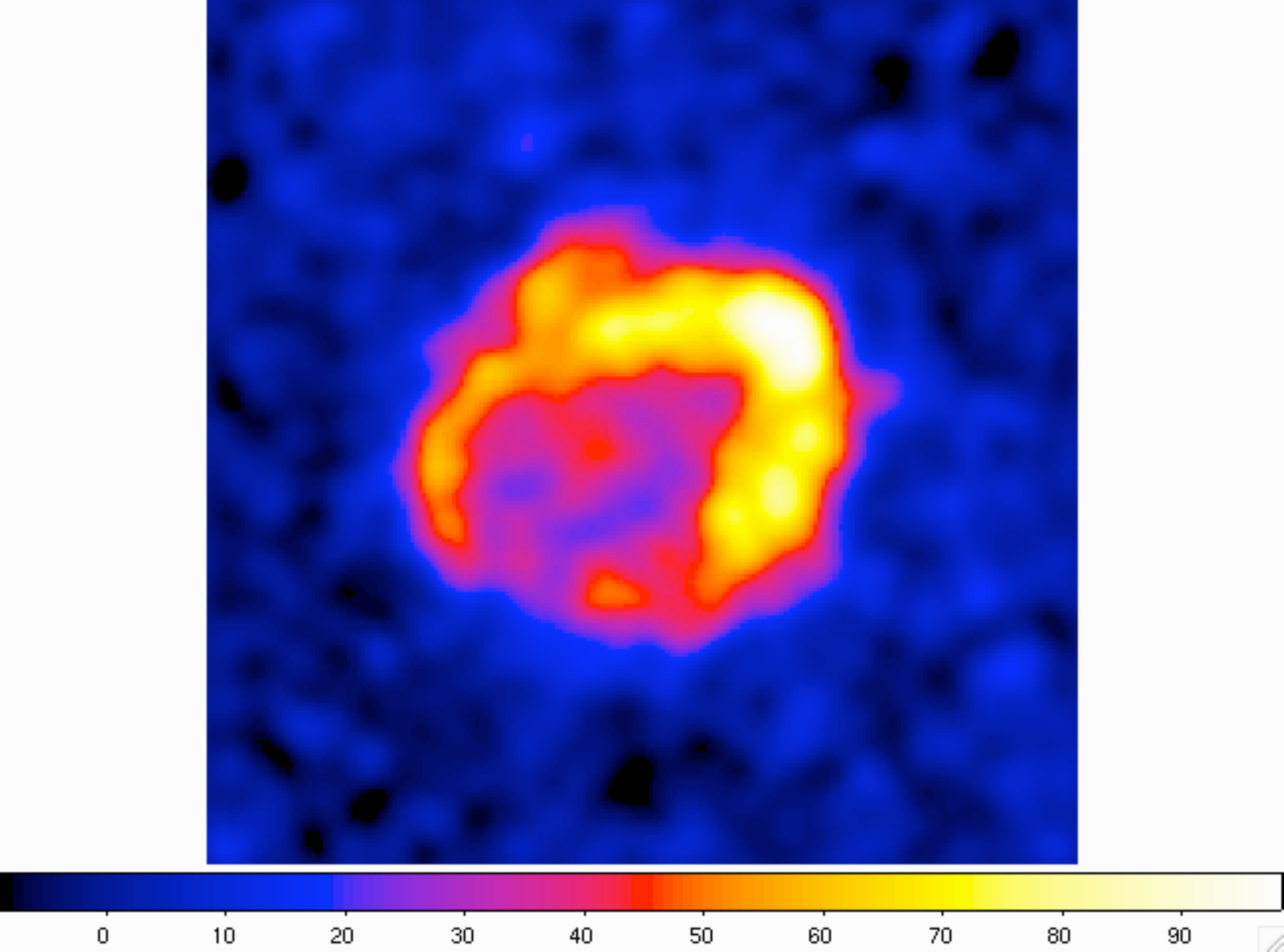}
\caption[Acceptance corrected gamma-ray excess image of the SNR, RX J1713-3946, observed by \HESS.]
{\label{fig:HESS-RXJ1713}Acceptance corrected gamma-ray excess image of the SNR, RX J1713-3946, observed by \HESS.}
}
\end{tabular}
\end{figure}

Imaging Atmospheric \Cherenkov\ Telescopes (IACTs) such as CANGAROO, \HESS, MAGIC and VERITAS detect the \Cherenkov\ light emitted by air showers induced by gamma rays (and cosmic-ray backgrounds) in 0.1--100~TeV energy range with a collection area of $\sim$2~\kmsq.
The shape of the the air shower retains the original direction of the incident gamma ray.
The number of \Cherenkov\ photons can be used to measure the energy of the shower.
One of the major challenges for the IACTs is reduction of cosmic-ray backgrounds which are $\sim$10,000 times more abundant than the gamma rays.
The shape of the shower can be used to reject the majority of cosmic rays by a factor of $\sim$1,000 since hadronic showers are more clumpy.
Stereoscopic observation by utilizing two or more telescopes separated by $\sim$100~m (as illustrated in Fig.~\ref{fig:IACT-stereo}) enables substantial reduction of the effective energy threshold, improvement of the angular and energy resolution for individual gamma rays, and further suppression of the cosmic-ray backgrounds.

\HESS\ and CANGAROO have been operating with 4 telescopes for three years.
VERITAS is expected to start 4-telescope operation in early 2007 and MAGIC will start 2-telescope operation in late 2007.
The \Cherenkov\ light is collected and focused by 10~m (CANGAROO), 12~m (VERITAS), 13~m (\HESS) and 17~m (MAGIC) mirrors and viewed by 500--1000 PMTs covering 3--5\degree\ field-of-view.

\begin{figure}
\centering
\begin{tabular}{ccc}
\parbox{7.5cm}{\centering
\includegraphics[height=6cm]{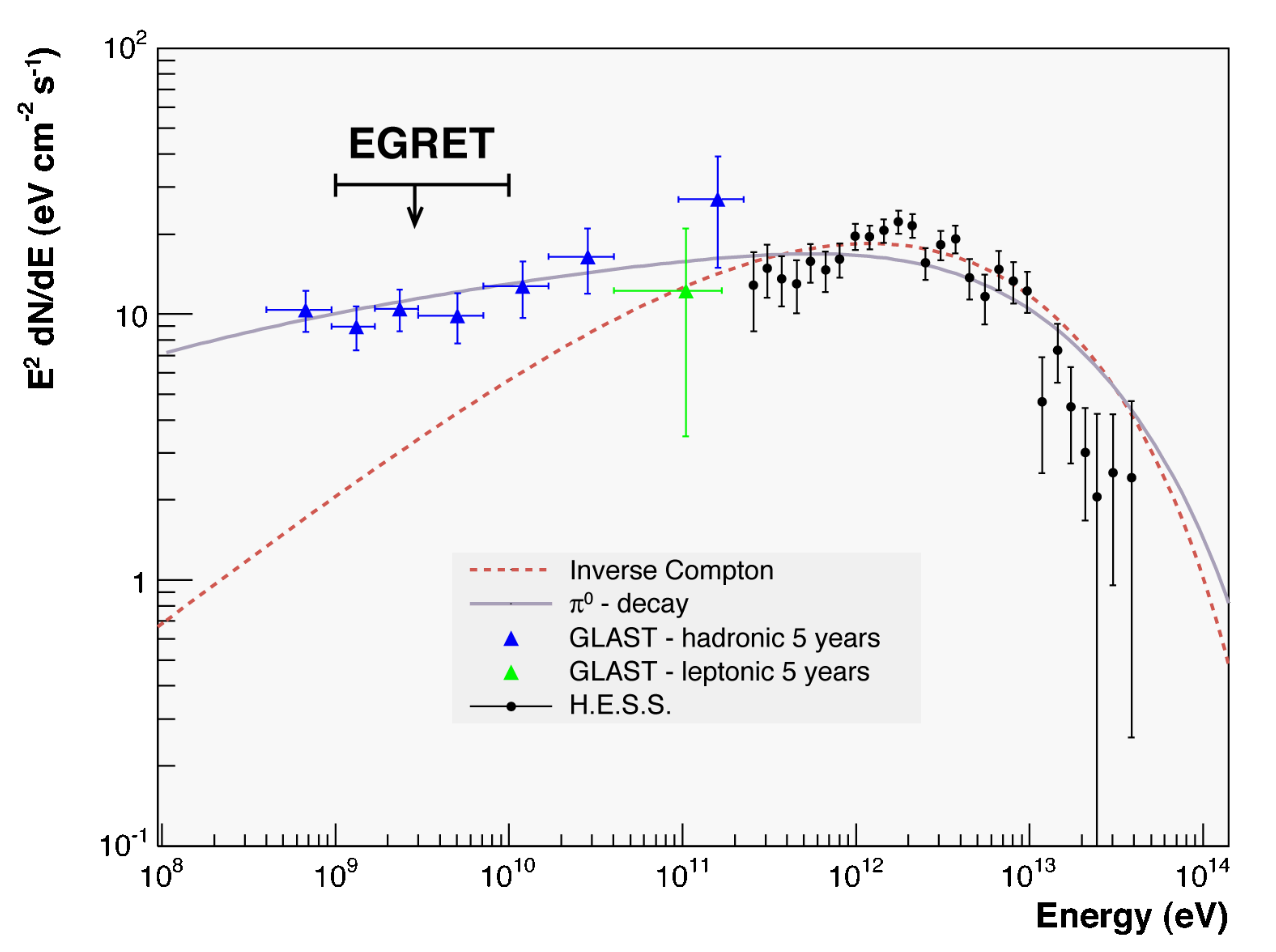}
\caption[Gamma-ray spectrum of RX J1713-3946 observed by \HESS. Simulated GLAST measurements for hadronic and electronic models are also shown.]
{\label{fig:SED-RXJ1713}Gamma-ray spectrum of RX J1713-3946 observed by \HESS. Simulated GLAST measurements for hadronic and electronic models are also shown.}
}
& &
\parbox{8.5cm}{\centering
\includegraphics[height=6cm]{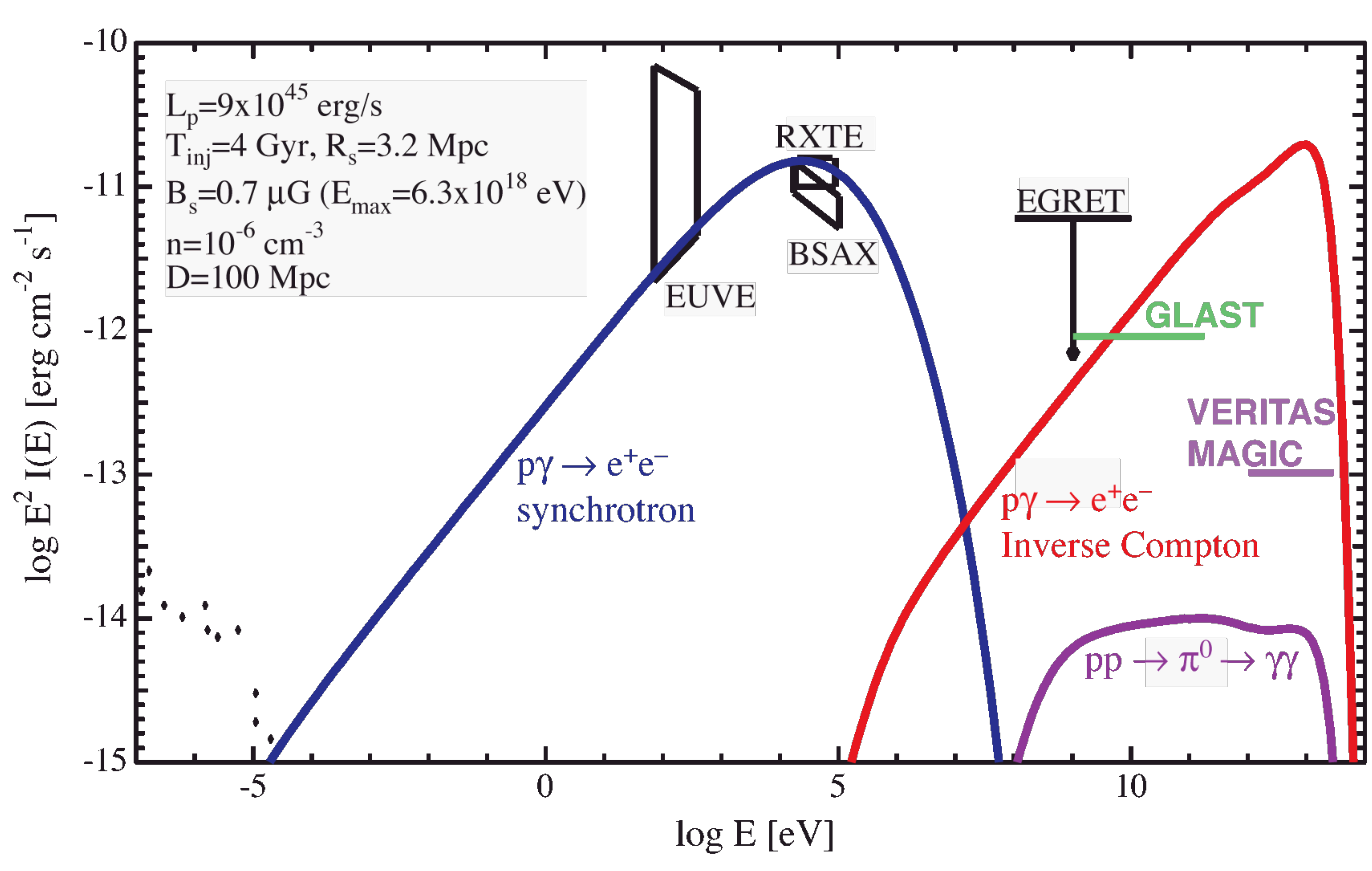}
\caption[Model prediction of gamma-ray spectra for a merging galaxy cluster, Coma. Expected sensitivity limits of GLAST and IACTs are also displayed.]
{\label{fig:Galaxy-Cluster-SED}Model prediction of gamma-ray spectra for a merging galaxy cluster, Coma. Expected sensitivity limits of GLAST and IACTs are also displayed.}
}
\end{tabular}
\end{figure}

The angular resolution of \HESS\ is 0.1\degree, which is comparable to that of the ASCA X-ray satellite.
Combined with the large field of view (5\degree), \HESS\ can resolve the morphology of extended gamma-ray sources.
Fig.~\ref{fig:HESS-RXJ1713} shows the latest TeV gamma-ray image\cite{HESS-RXJ1713} of a shell type SNR, RX J1713-3946, which was the first SNR imaged in a TeV gamma-ray band by \HESS\cite{Aharonian04}.
This observation presents undisputed evidence that electrons or protons were accelerated to at least $10^{14}$~eV, which is very close to the ``knee" energies.
Morphological similarity with X-ray images of the same SNR may suggest that Compton scattering of CMB photons by high energy electrons and positrons is responsible for TeV gamma-ray emissions since the X-ray is considered to originate from the synchrotron radiation of the same lepton population.
However, spectral features of the \HESS\ observation do not necessarily agree with the prediction by such leptonic models.
The Large Area Telescope (LAT) instrument of the Gamma-ray Large-Area Space Telescope (GLAST) mission described below can provide conclusive evidences on this matter since the spectral predictions by hadronic and leptonic models diverge below 100~GeV as presented in Fig.~\ref{fig:SED-RXJ1713}\cite{Funk}.
It shows the \HESS\ measurement result along with simulated GLAST measurements for the hadronic and leptonic cases.

Merging galaxy clusters are considered to be among the candidates for the origin of UHECRs due to the sheer size of their acceleration volume.
UHECRs accelerated in such clusters can interact with CMB photons producing UHE electrons and positrons. (It is impossible to accelerate leptons to such high energies due to synchrotron loss.)
These leptons can produce hard X-rays via synchrotron radiation and gamma rays via Compton scattering of CMB photons.
Multi-wavelength observation of merging galaxy clusters in the X-ray and gamma-ray bands can constrain such models as demonstrated in Fig.~\ref{fig:Galaxy-Cluster-SED}\cite{Inoue}.
It shows the model prediction for X-ray and gamma-ray spectra from synchrotron and Compton processes constrained by existing X-ray observations of the Coma cluster.
Observation of the Coma galaxy cluster by IACTs in the northern hemisphere such as MAGIC or VERITAS will very likely provide a conclusive answer on this scenario.

\section{GLAST}
\begin{figure}
\centering
\begin{tabular}{cc}
\parbox{8cm}{\centering
\includegraphics[height=6cm]{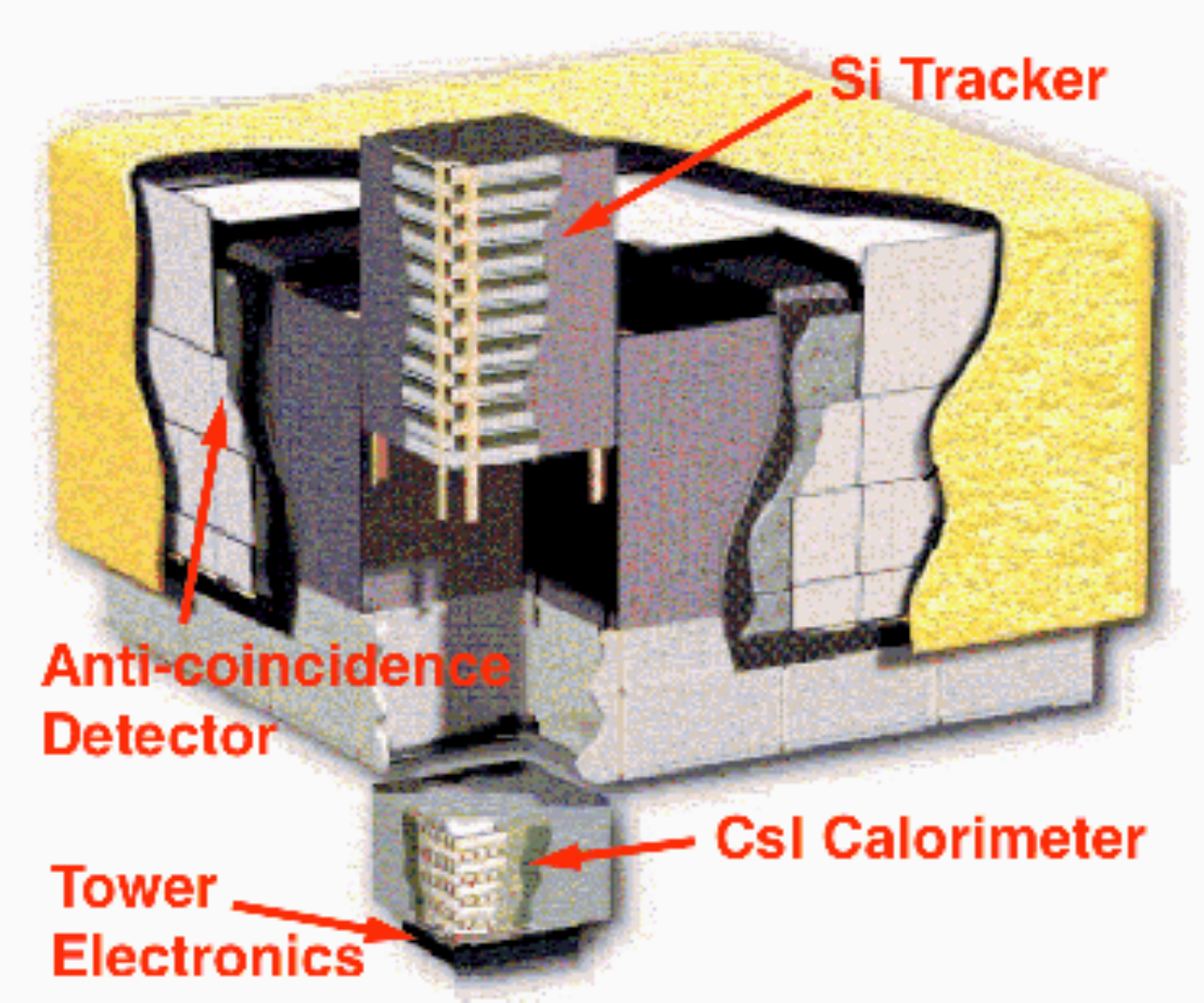}
\caption[Schematic view of GLAST/LAT instrument.]
{\label{fig:GLAST-schematic}Schematic view of GLAST/LAT instrument.}
}
&
\parbox{8.4cm}{\centering   
   \includegraphics[height=6cm]{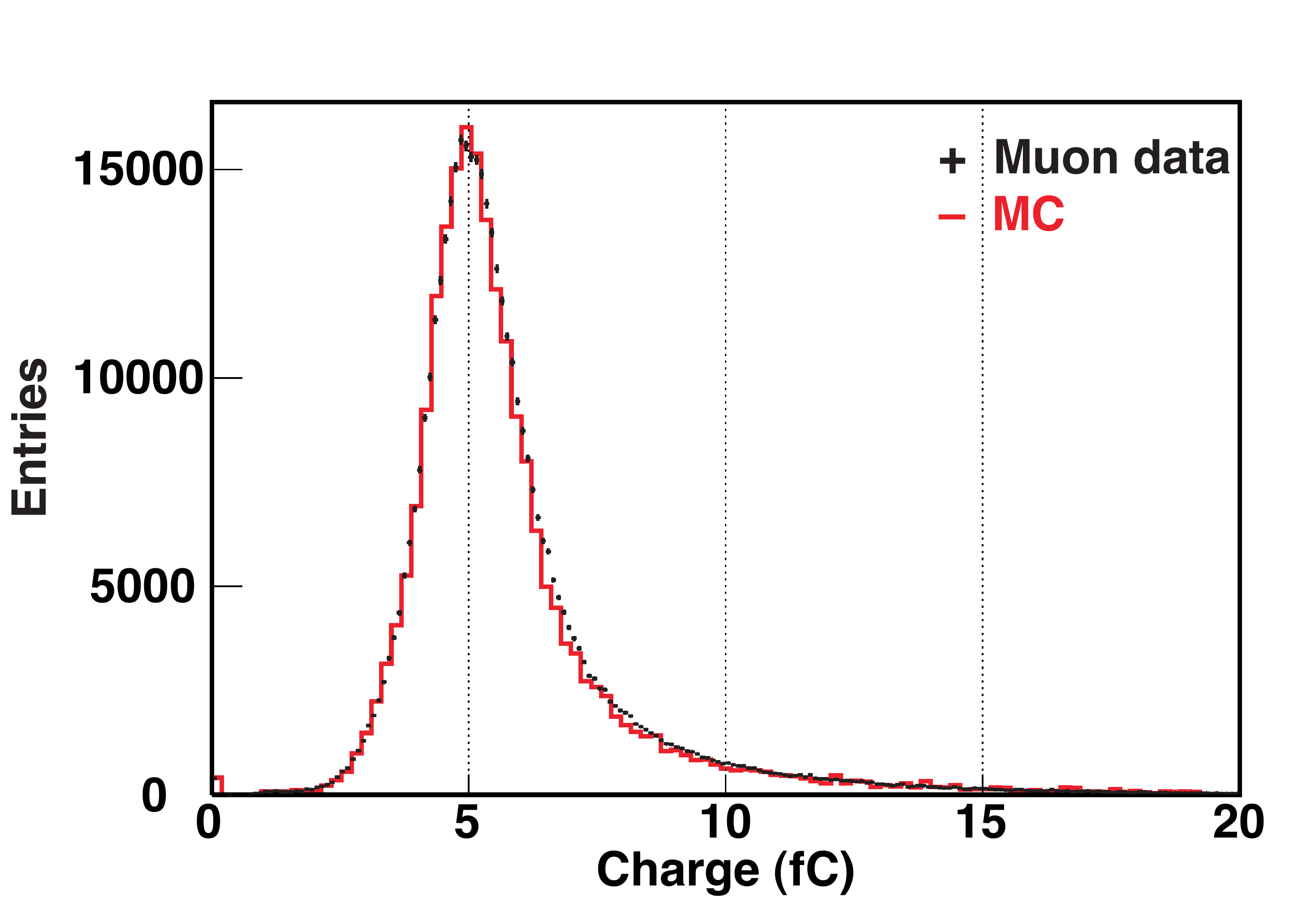}
   \caption{ \label{fig:GLAST-tot} Comparison of the TOT distribution of cosmic-ray muon tracks for data (points) and a Monte Carlo simulation sample (histogram). }
   }
   \end{tabular}
   \end{figure}

The GLAST/LAT instrument\cite{GLAST} is a pair-conversion gamma-ray detector similar in concept to the previous NASA high-energy gamma-ray instrument EGRET on the Compton Gamma-Ray Observatory\cite{EGRET}.
Fig.~\ref{fig:GLAST-schematic} shows a schematic view of the GLAST/LAT instrument.
The instrument has an active area of $>1.9$~\msq\ with a weight less than 3000~kg and a power consumption below 650~W.
High-energy (20~MeV--300~GeV) gamma-rays convert into electron-positron pairs in one of 16 layers of tungsten foils.
The charged particles pass through up to 36 layers of position-sensitive detectors interleaved with the tungsten, the ``tracker," leaving behind tracks pointing back toward the origin of the gamma ray.  
After passing through the last tracking layer they enter a calorimeter composed of bars of cesium-iodide crystals read out by PIN diodes.
The calorimeter furnishes the energy measurement of the incident gamma ray.
Since the radiation length of the calorimeter is 8.4, it utilizes the the shower shape to estimate the energy at the high end of the energy range.
A third detector system, the anti-coincidence detector (ACD), surrounds the top and sides of the tracking instrument.
It consists of panels of plastic scintillator read out by wave-shifting fibers and photo-multiplier tubes and is used to veto charged cosmic-ray events such as electrons, protons or heavier nuclei with 99.97\% efficiency.
Each scintillator tile is readout by two PMTs to avoid single point of failure.

In the LAT, the tracker and calorimeter are segmented into 16 ``towers," which are covered by the ACD and a thermal blanket and meteor shield.
An aluminum grid supports the detector modules and the data acquisition system and computers, which are located below the calorimeter modules.
The LAT is designed to improve upon EGRET's sensitivity to astrophysical gamma-ray sources by well over a factor of 10.
That is accomplished partly by sheer size, but also by use of the silicon-strip detectors used in the tracker system.

Each of the 16 tracker modules is composed of a stack of 19 ``trays."
A tray is a stiff, lightweight carbon-composite panel with silicon-strip detectors (SSDs) bonded on both sides, with the strips on the top parallel to those on the bottom.
Also bonded to the bottom surface of all but the 3 lowest trays, between the panel and the detectors, is an array of tungsten foils, one to match the active area of each detector wafer.
The thickness of the tungsten foil is 3\% radiation length for the upper 12 trays (light-converter trays), 18\% radiation length for the next 4 trays (thick-converter trays). 
The last 3 trays do not have tungsten foils.
Each tray is rotated 90\degree\ with respect to the one above or below.
The detectors on the bottom of a tray combine with those on the top of the tray below to form a 90\degree\ stereo x,y pair with a 2~mm gap between them, and with the tungsten converter foils located just above.

Each front-end electronics multi-chip module (MCM) supports the readout of 1536 silicon strips.
It consists of a single printed wiring board (PWB) upon which are mounted 24 64-channel amplifier-discriminator ASICs (GTFE), two digital readout-controller ASICs (GTRC), the right-angle interconnect, bias and termination resistors, decoupling capacitors, resettable fuses, and two nano-connectors.
Each nano-connector plugs into a long flex-circuit cable, each of which interfaces 9 MCMs to the data-acquisition electronics located below the calorimeter in the Tower Electronics Module (TEM).
Thus on each of the 4 sides of a tracker module one finds 9 readout boards to support 9 layers of silicon-strip detectors, which send their data to the TEM via two flex-circuit cables.
Each channel in the GTFE has a preamplifier, shaping amplifier, and discriminator\cite{Johnson98}.  
The amplified detector signals are discriminated by a single threshold per GTFE chip; no other measurement of the signal size is made within the GTFE.
The GTFE chips are arranged on the MCM in 4 groups of 6.
Each group reads out one SSD ladder, which consists of 4 SSDs connected in series to yield strips of about 36~cm effective length.
Each GTFE chip has two command decoders, one that listens to the left-hand GTRC, and a second that listens to the right-hand GTRC.
Each GTFE also has two output data shift registers, one that moves data to the left, and a second that moves data to the right.
Trigger information is formed within each GTFE chip from a logical OR of the 64 channels, of which any arbitrary set can be masked.
The OR signal is passed to the left or right, depending on the setting of the chip, and combined with the OR of the neighbor, and so on down the line, until the GTRC receives a logical OR of all non-masked channels in those chips that it controls.
A counter in the GTRC measures the length of the layer-OR signal (time-over-threshold, TOT) and buffers the result for inclusion in the event data stream.

The GLAST/LAT instrument has been assembled and tested against vibrations and thermal cycles in the vacuum.
The instrument was integrated with a spacecraft in December 2006 and is expected to be launched in early 2008.

The performance of the instrument has been checked out using cosmic-ray muons.
The average hit efficiency is found to be 99.6\%, and 99\% of the layers have more than 98\% hit efficiency.
The discriminator threshold is nominally set at 1.4~fC (approximately 1/4 of a minimum ionizing particle signal for 400~$\mu$m thick Si.)
Due to channel gain variation, the threshold dispersion within each GTFE is 5.2\%, while the threshold dispersion between GTFEs after calibration is 2.7\%, due to the granularity of the threshold DAC.
The trigger time walk needs to be minimized to suppress chance coincidences with the ACD veto.
After the threshold DACs were calibrated, the trigger time walk between injection charges of 2.5~fC and 20--30~fC was measured.
It is found that 99.8\% of the channels are within the 0.6~$\mu$s requirement.
The TOT provides a maximum pulse height measurement for each GTRC (a half layer).
The TOT gain varies from channel to channel by 30\% rms due to the dispersion of the shaper fall time. (The peaking time is relatively uniform.)
The relative gain of the TOT is calibrated using test pulses.
Absolute TOT calibration for each GTFE is performed using cosmic-ray muon tracks, which also set the absolute scale for the GTFE charge injector.
Fig.~\ref{fig:GLAST-tot} compares the TOT distribution of cosmic-ray muon tracks for data after calibration (points) and a Monte Carlo simulation sample (histogram).
The distributions agree with each other very well.

Beam tests of spare modules were carried out at CERN using tagged-photon, electron and proton beam to characterize the instrument response.
Monte Carlo simulation will be tuned to reproduce the beam test result so that the LAT performance can be simulated more precisely.
Electron and proton beams are also used to understand the instrument response to cosmic-ray backgrounds.

\section{Constraints on Satellite Instruments}
In addition to apparent constraints on satellite-borne or balloon-borne instruments, such as power, volume and mass, there are numerous constraints or requirements to achieve high reliabilities as listed below.
\begin{itemize}
\item It is critical to avoid single points of failure in satellite instruments for obvious reasons. 
Redundant readout paths and modular design are useful solutions for this problem.
\item Low power is important not only because of power source constraint, but also because of heat transfer.
In Space, lack of air makes it very difficult to conduct heat out of heat sources.
Custom ASICs as well as low-speed operation is required to achieve this.
\item NASA requires Space qualified parts, which does not necessarily mean high quality since NASA qualifications heavily depend on screening and traceable records.
Commercial parts could be of higher quality since low failure rate is crucial to achieve high volume.
Space qualified parts could have very long lead time (12--26 weeks) for delivery and have very bad cost/perfromance ratios.
For example, a 0.8~GHz PowerPC single board computer costs 1/4 million dollars and a 2,000 flip-flop FPGA costs 4,000 dollars.
\end{itemize}
Details of electronics for satellite experiments can be found in ref \cite{Johnson06}.


\section{Conclusions}
The next generation of detectors to study the origin of cosmic rays, UHECRs and cosmic antiparticles have been or are being constructed.
These detectors on ground and in Space deal with a variety of messengers (charged particles, neutrinos and gamma rays) over a wide energy range, $10^7$--$10^{20}$~eV.
The Auger Observatory, IceCube and ANITA will provide definitive answer on GZK suppression and may find UHECR sources.
PAMELA will provide precise measurements of the antiparticle flux with stringent implications on dark matter properties.
Gamma-ray detectors (IACTs and GLAST) will provide conclusive evidences on whether SNRs are cosmic-ray accelerators.
This is definitely a very exciting time and we are at the beginning a new era of cosmic-ray astrophysics.





\begin{thebibliography}{99}
\bibitem{Blandford87} R. Blandford and D. Eichler, Phys. Rep., {\bf 154}, 1 (1987).
\bibitem{Ginzburg64} V.L. Ginzburg and S.I. Syrovatskii, ``Origin of Cosmic Rays", Macmillan, 1964.
\bibitem{Koyama95} K. Koyama, R. Petre, E. V. Gotthelf, U. Hwang, M. Matsuura, M. Ozaki and S. S. Holt,  Nature, {\bf 378}, 255 (1995).
\bibitem{Enomoto02} R. Enomoto \etal\ (CANGAROO Collaboration) Nature {\bf 416}, 823 (2002).
\bibitem{Aharonian04} F.~Aharonian \etal\ (\HESS\ Collaboration) Nature, {\bf 432}, 75 (2004)
\bibitem{GZK} K. Greisen, Phys. Rev. Lett. {\bf 16}, 748 (1966), T Zatsepin and V. A. Kuzmin, JETP. Lett. {\bf 4}, 78 (1966).
\bibitem{AGASA} M. Takeda \etal\ Astropart. Phys. {\bf 19}, 447 (2003).
\bibitem{HiRes} R.U. Abbasi \etal\ Phys. Rev. Lett. {\bf 92}, 151101 (2006).
\bibitem{PAO} J. Abraham \etal\ (Pierre Auger Collaboration), Nucl. Inst. Meth. A. {\bf 523}, 50 (2004); 
Auger Collaboration, Proceedings of 29th ICRC, Pune, India (2005), arg-bertou-X-abs1-he14-oral; usa-mostafa-M-abs1-he14-oral; usa-allison-PS-abs1-he14-poster; usa-bauleo-PM-abs1-he14-poster; mex-nellen-L-abs1-he14-oral; bra-bonifazi-C-abs1-he14-oral; ita-ghia-P-abs1-he14-oral.
\bibitem{PAO-result} Paul Mantsch (Pierre Auger Collaboration), astro-ph/0604114.
\bibitem{PAMELA} P. Picozza \etal\ (PAMELA collaboration), astro-ph/0608697.
\bibitem{IceCube} Albrecht Karle (IceCube collaboration), Nucl. Inst. and Meth. A {\bf 567}, 438 (2006); Paolo Desiati (IceCube collaboration), astro-ph/0611603.
\bibitem{Ahrens} J.~Ahrens \etal\ (IceCube collaboration), Astropart. Phys. {\bf 20} 507 (2004).
\bibitem{MPR} K. Mannheim, R.J. Protheroe, J.P. Rachen, Phys. Rev. D
{\bf 63} 023003 (2001).
\bibitem{SS} F.W. Stecker, M.H. Salamon, Space Sci. Rev. {\bf 75} 341 (1996).
\bibitem{WB} E. Waxman, J. Bahcall, Phys. Rev. D {\bf 59} (1999) 023002.
\bibitem{Askaryan} G.A. Askaryan, Sov. Phys. JETP {\bf 14}, 441 (1962);
{\bf 21}, 658 (1965).
\bibitem{ANITA} P. Mio\v{c}inovi\'{c} (ANITA collaboration), astro-ph/0503304; G.~.S~.Varner \etal, SLAC-PUB-11872 Presented at 9th Int. SNIC Symposim, Stanford, California, 2006.
\bibitem{LAB3} G. Varner, J. Cao, M. Wilcox and P. Gorham, physics/0509023.
\bibitem{ANITA-beam-test} P. W. Gorham \etal\ (ANITA collaboration), astro-ph/0611008.
\bibitem{HESS-RXJ1713} F.~Aharonian \etal\ (\HESS\ Collaboration) Astron. Astrophys. {\bf 464} 235 (2007).
\bibitem{Funk} provided by Stefan Funk of GLAST collaboration.
\bibitem{Inoue} S. Inoue \etal, Astroph. Jour. {\bf 628}, 9 (2005)
\bibitem{GLAST} W.B. Atwood, Nucl. Instrum. Meth. A {\bf 342}, 302 (1994).
N. Gehrels and P. Michelson, Astropart. Phys. {\bf 11}, 277 (1999).
\bibitem{EGRET}	D.J. Thompson, et al. Astroph. Jour. Suppl. {\bf 86}, 629 (1993).
\bibitem{Johnson98} R.~P.~Johnson, P.~Poplevin, H.~F.-W.~Sadrozinski, and E.~N.~Spencer, IEEE Trans. Nucl. Sci. {\bf 45}, 927 (1998).
\bibitem{Johnson06} R.~P.~Johnson, Presented at 9th Int. SNIC Symposim, Stanford, California, 2006.
\end{thebibliography}
\end{document}